\documentclass[sigplan,screen,nonacm]{acmart} 

\setcopyright{none}
\acmConference{Rutgers DCS Technical Report DCS-TR-754}
\acmYear{2021}
\acmDOI{}
\acmISBN{}

\usepackage{booktabs}   
\usepackage{subcaption} 
\usepackage{amsmath}
\usepackage[percent]{overpic}
\usepackage{xspace}
\usepackage{listings}
\usepackage{xcolor}
\usepackage{empheq}
\usepackage{pifont}
\usepackage[linesnumbered, vlined]{algorithm2e}

\newcommand{\rlibm}{\textsc{RLibm}\xspace}
\newcommand{\tool}{\textsc{RLibm-32}\xspace}

\newcommand{\eg}{\emph{e.g.}\xspace}
\newcommand{\ie}{\emph{i.e.}\xspace}   \newcommand{\cmark}{\ding{51}}%
\newcommand{\xmark}{\ding{55}}%

\makeatletter
\let\old@lstKV@SwitchCases\lstKV@SwitchCases
\def\lstKV@SwitchCases#1#2#3{}
\makeatother
\usepackage{lstlinebgrd}
\makeatletter
\let\lstKV@SwitchCases\old@lstKV@SwitchCases

\definecolor{lbcolor}{rgb}{0.875,0.875,0.875}
\begin{document}

\title[]{\textsc{RLIBM-32}: High Performance Correctly Rounded Math Libraries for 32-bit
  Floating Point Representations \\ \vspace{10pt} \small{Rutgers Department of Computer Science Technical Report DCS-TR-754 \\ Extended version of the PLDI 2021 paper}}




\author{Jay P. Lim}
\orcid{0000-0002-7572-4017}             
\affiliation{
  \department{Department of Computer Science}              
  \institution{Rutgers University}            
  \country{United States}                    
}
\email{jpl169@cs.rutgers.edu}          

\author{Santosh Nagarakatte}
\orcid{0000-0002-5048-8548}             
\affiliation{
  \department{Department of Computer Science}              
  \institution{Rutgers University}            
  \country{United States}                    
}
\email{santosh.nagarakatte@cs.rutgers.edu}          


\begin{abstract}
  This paper proposes a set of techniques to develop correctly rounded
  math libraries for 32-bit float and posit types.  It enhances our
  \rlibm approach that frames the problem of generating correctly
  rounded libraries as a linear programming problem in the context of
  16-bit types to scale to 32-bit types. Specifically, this paper
  proposes new algorithms to (1) generate polynomials that produce
  correctly rounded outputs for all inputs using counterexample guided
  polynomial generation, (2) generate efficient piecewise polynomials
  with bit-pattern based domain splitting, and (3) deduce the amount
  of freedom available to produce correct results when range reduction
  involves multiple elementary functions. The resultant math library
  for the 32-bit float type is faster than state-of-the-art math
  libraries while producing the correct output for all inputs. We have
  also developed a set of correctly rounded elementary functions for
  32-bit posits.
\end{abstract}



\maketitle

\section{Introduction}
Math libraries provide implementations of elementary functions (\eg
$log(x)$, $e^x$, $cos(x)$)~\cite{Muller:elemfunc:book:2005}. They are crucial
components in various domains ranging from scientific computing to
machine learning. Designing math libraries is a challenging task
because they are expected to provide correct results for all inputs
and also have high performance. These elementary functions are
typically approximated with some hardware supported representation for
performance.

Given a representation~$\mathbb{T}$ with finite precision (\eg,
\texttt{float}), the correctly rounded result of an elementary
function $f$ for an input $x \in \mathbb{T}$ is defined as the value
of $f(x)$ computed with real numbers and then rounded to a value in
the representation~$\mathbb{T}$. The IEEE-754 standard recommends the
generation of correctly rounded results for elementary functions.
Seminal prior work on generating approximations for elementary
functions has resulted in numerous implementations that have reduced
error significantly~\cite{Lefevre:toward:tc:1998,
  Chevillard:sollya:icms:2010, Brisebarre:maceffi:toms:2006,
  Chevillard:infnorm:qsic:2007, Chevillard:ub:tcs:2011,
  Olga:metalibm:icms:2014, Brunie:metalibm:ca:2015,
  Jeannerod:sqrt:tc:2011, Bui:exp:ccece:1999,
  Abraham:fastcorrect:toms:1991, 
  Fousse:toms:2007:mpfr}. Further, numerous
correctly rounded libraries have also been
developed~\cite{Daramy:crlibm:doc,
  Fousse:toms:2007:mpfr}. Unfortunately, they
are not widely used due to performance considerations. Moreover,
widely used libraries do not produce correct results for all inputs.

\textbf{Mini-max approaches.} Most prior approaches identify a
polynomial that minimizes the maximum error among all input points
(\ie, a mini-max approach) compared to the real value of the
elementary function using the Weierstrass approximation theorem and
the Chebyshev alternation
theorem~\cite{Trefethen:chebyshev:book:2012}. The Weierstrass
approximation theorem states that if $f$ is a continuous real-valued
function on $[a,b]$ and $\epsilon > 0$, there exists a polynomial $P$
such that $|f(x) - P(x)| < \epsilon$ for all $x \in [a,b]$. The
Chebyshev alternation theorem provides the condition for such a
polynomial: a polynomial of degree $d$ that minimizes the maximum
error will have at least $d+2$ points where it has the absolute
maximum error and the error alternates in sign. Remez
algorithm~\cite{Remes:algorithm:1934,Muller:elemfunc:book:2005} is a
procedure to identify such mini-max polynomials. The maximum
approximation error has to be below the error threshold required to
produce correct results for all inputs.

As approximating a polynomial in a small domain $[a,b]$ is much
easier, the input domain of the function is reduced using range
reduction~\cite{Cody:book:1980,Tang:log:toms:1990,Lim:rlibm:arxiv:2020}. The
approximated result is adjusted to produce the result for the original
input (\ie, output compensation).  Both range reduction and polynomial
evaluation in a representation with finite precision will have some
numerical errors. The combination of approximation errors with the
mini-max approach and numerical errors with polynomial evaluation,
range reduction, and output compensation can result in wrong results.

\begin{figure*}
  \centerline{\includegraphics[width=\linewidth]{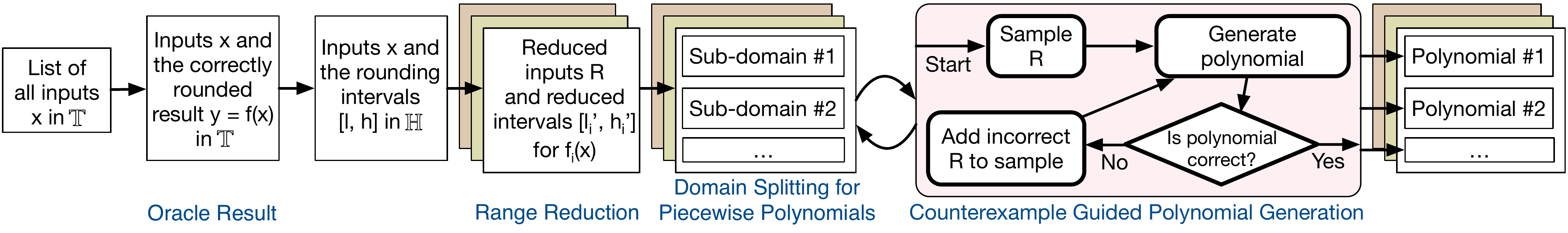}}
  \caption{Steps in our approach to generate correctly rounded
    libraries for 32-bit types ($\mathbb{T}$).}
    \label{fig:workflow}
\end{figure*}
\textbf{\rlibm.} Our \rlibm approach~\cite{lim:rlibm:popl:2021,
  Lim:rlibm:arxiv:2020} generates polynomials that approximate the
correctly rounded result rather than the real value of the elementary
function. The generation of the polynomial considers errors in
polynomial approximation and numerical errors in polynomial
evaluation, range reduction, and output compensation to produce the
correctly rounded output for all inputs. The task of generating the
polynomial is then structured as a linear programming (LP)
problem. The \rlibm approach first computes the correctly rounded
result for each input in a target representation $\mathbb{T}$ using an
oracle (\eg, the MPFR library~\cite{Fousse:toms:2007:mpfr}). Given the
correctly rounded result for an input, it finds an interval in
\texttt{double} precision such that every value in the interval rounds
to the correctly rounded result, which is called the rounding
interval.  The rounding intervals are further constrained to account
for numerical errors during range reduction and output compensation.
Subsequently, it attempts to generate a polynomial of degree $d$ using
an LP solver, which when evaluated with an input produces a result
that lies within the rounding interval. Using the \rlibm approach, we
have been successful in generating correctly rounded libraries with
16-bit types such as \texttt{bfloat16} and \texttt{posit16}.

\textbf{Challenges in scaling to 32-bits.} To extend our \rlibm
approach to 32-bit data types, we have to address the following
challenges. First, modern LP solvers can handle a few thousand
constraints.  A naive use of the \rlibm approach with 32-bit types
will generate more than a billion constraints, which is beyond the
capabilities of current LP solvers. Second, it may not be feasible to
generate a single polynomial of a reasonable degree given the large
number of constraints. Third, LP solvers are sensitive to the
condition number of the system of constraints.  LP solvers will not be
able to solve an ill-conditioned system of constraints.  An effective
range reduction is a strategy to address it. Although there are
excellent books on range reduction~\cite{Cody:book:1980}, these
techniques need to be adapted to work with our \rlibm
approach. Fourth, some range reduction strategies need multiple
elementary functions themselves (\eg, $sinpi(x)$). Finally, we need to
ensure that output compensation does not experience pathological
cancellation errors (\eg, $cospi(x)$).

\textbf{This paper.} Our goal is to generate efficient implementations
of elementary functions that produce correctly rounded results for all
inputs with 32-bit types. This paper extends our \rlibm approach to
scale to 32-bit FP types to address the challenges described above. We
propose (1) sampling of inputs with counterexample guided polynomial
generation to handle the large input space, (2) generation of
piecewise polynomials for efficiency, (3) deduction of rounding
intervals when a range reduction technique uses multiple elementary
functions, and (4) modified range reduction techniques for some
elementary functions to address cancellation errors in output
compensation. Figure~\ref{fig:workflow} pictorially represents our
approach to scale to 32-bit data types.

\textbf{Counterexample guided polynomial generation.} We sample inputs
proportional to the number of representable values in a given input
domain $[a,b]$ with a 32-bit representation~$\mathbb{T}$.  To generate
polynomials that produce the correctly rounded result for every input,
it is not necessary to consider every input and its rounding
interval. We primarily need to consider those rounding intervals that
are highly constrained.
For each input in the sample, we generate the oracle result using the
MPFR library. We compute the rounding interval in double precision
(\ie, set of values in the \texttt{double} type that round to the
oracle result). We generate LP constraints to create a polynomial of
degree $d$ such that it evaluates to a value in the rounding interval
for each input in the sample.
If the initial sample generates a polynomial that produces the
correctly rounded output for all values in $[a,b]$, then the process
terminates. Otherwise, we add counterexamples to the sample and repeat
the process. The size of the sample is bounded by the number of
constraints that the LP solver can process.

\textbf{Piecewise polynomials.} When either the number of inputs in
the sample exceeds our LP constraint threshold or the LP solver is not
able to generate a polynomial, we split the input domain $[a,b]$ to
$[a, b')$ and $[b', b]$ to generate piecewise polynomials using the
  above process for each input sub-domain. We choose the splitting
  point such that we can identify the sub-domain quickly using a few
  bits of the input, which results in efficient implementations. The
  ability to generate piecewise polynomials ensures that our resultant
  polynomials are of a lower degree and provide performance
  improvements when compared to state-of-the-art libraries.

\textbf{Range reduction with multiple functions.}  We propose new
algorithms to deduce rounding intervals for a class of range reduction
techniques that involve multiple elementary functions.  Range
reduction reduces the input $x$ to $x'$. The creation of the
polynomial happens with the reduced inputs. The output of the
polynomial $P(x')$ should be adjusted to compute the correctly rounded
result for $x$, which is called output compensation. We have to deduce
the rounding intervals for the reduced input $x'$ that considers the
numerical error in range reduction, polynomial evaluation, and output
compensation. We propose new techniques to create reduced rounding
intervals when range reduction uses multiple elementary functions
(\eg, $sinpi(x)$ in Section~\ref{sec:overview}). These techniques
allow us to perform range reduction on functions that otherwise cause
condition number issues with the LP formulation~(\ie, $sinh(x)$ or
$cosh(x)$). Further, we develop modified range reduction techniques
for some elementary functions to avoid cancellation errors in output
compensation (\eg, $cospi(x)$ in Section~\ref{sec:casestudy}).

\textbf{The \textsc{Rlibm-32} prototype.} We have developed library
generators and correctly rounded libraries for multiple 32-bit data
types: IEEE-754 float and posits. Our elementary functions for
\texttt{floats} are faster than existing libraries: Intel's libm,
Glibc's libm, CR-LIBM~\cite{Daramy:crlibm:doc}, and
Metalibm~\cite{Olga:metalibm:icms:2014}. Unlike existing libraries,
our functions produce correctly rounded results for all inputs. We
have developed the first correctly rounded implementations of
functions for 32-bit posits.

\section{Overview of Our Approach with $sinpi(x)$}
\label{sec:overview}
We provide an overview of our approach for generating piecewise
polynomials for $sinpi(x)$ (\ie, $sin(\pi x)$) with a 32-bit
float. The function $sinpi(x)$ is defined for $x \in (-\infty,
\infty)$. There are four billion inputs with a 32-bit float. There are
three kinds of special cases:

\begin{small}
\[
  sinpi(x) =
  \begin{cases}
    \pi x & \text{if } |x| < 1.173\dots \times 10^{-7} \\
    0 & \text{if } |x| >=  2^{23}\\
    NaN & \text{if } x = NaN \text{ or } x = \pm \infty
  \end{cases}
  \]
\end{small}

For the first class of special cases, we compute $\pi x$ in double and
round the result to float, which produces the correctly rounded result
for those inputs.

\subsection{Our Range Reduction for $sinpi(x)$}
\label{overview:rangereduction}
After considering special cases, there are close to 800 million float
inputs that need to be approximated with polynomials. Using \rlibm's
approach directly with an LP solver will fail. Next, we perform range
reduction to reduce the domain for polynomial approximation.  The key
idea is to use periodicity and trigonometric identities of
$sinpi(x)$. We transform input $x$ into $x = 2.0 \times I + J$, where
$I$ is an integer and $J \in [0, 2)$. As a result of periodicity,
  $sinpi(x) = sinpi(J)$.
Next, we further split $J$ into $J = K + L$ where $K \in \{0, 1\}$ is
the integral part of $J$ and $L \in [0, 1)$ is the fractional
  part. Then, $sinpi(J)$ can be computed as,

\begin{small}
\[
  sinpi(J) = (-1)^{K} \times sinpi(L)
\]
\end{small}

Given that $sinpi$ between $[0.5, 1)$ is a mirror image of values
  between $[0, 0.5)$, we further reduce as follows:
\begin{small}
\[
  L' =
  \begin{cases}
    L & \text{if } L \leq 0.5 \\
    1.0 - L & \text{if } L > 0.5
  \end{cases}
\]
\end{small}

From Sterbenz lemma~\cite{Sterbenz:floating:book:1974}, the expression
$1.0 - L$ can be computed exactly.  Hence, $sinpi(L) =
sinpi(L')$. Even after reducing the input $x$ to $L' \in [0, 0.5]$,
there are around 184 million inputs with a 32-bit float in this
reduced domain.

To enable easier polynomial approximation, we further reduce $L'$ to a
value between $[0, \frac{1}{512}]$. We split $L'$ as $L' =
\frac{N}{512} + R$ where $N$ is an integer in the set $\{0, 1, \dots,
  255\}$ and $R$ is a fraction that lies in $[0, \frac{1}{512}]$. There
are $110$ million reduced inputs in $R$ ignoring special cases.
Now, $sinpi(L')$ can be computed using the trigonometric identity
$sinpi(a + b) = sinpi(a)cospi(b) + cospi(a)sinpi(b)$ as follows,

\begin{small}
\[
  sinpi(L') = sinpi(\frac{N}{512}) cospi(R) + cospi(\frac{N}{512})
  sinpi(R)
\]
\end{small}

We precompute the values for $sinpi(\frac{N}{512})$ and
$cospi(\frac{N}{512})$ in lookup tables (\ie, 512 values in total).
Finally, we approximate $sinpi(R)$ and $cospi(R)$ for the reduced
input domain $R \in [0, \frac{1}{512}]$.
\emph{To approximate $sinpi(x)$ for the entire domain, the range
reduction requires us to approximate $sinpi$ and $cospi$ over the
reduced domain $R$}. We can compute the result for $sinpi(x)$ as
follows, 

\begin{small}
\[
sinpi(x) = (-1)^K \times (sinpi(\frac{N}{512}) cospi(R) + cospi(\frac{N}{512})
  sinpi(R))
\]
\end{small}

\begin{figure*}%
  \small
  \includegraphics[width=\textwidth]{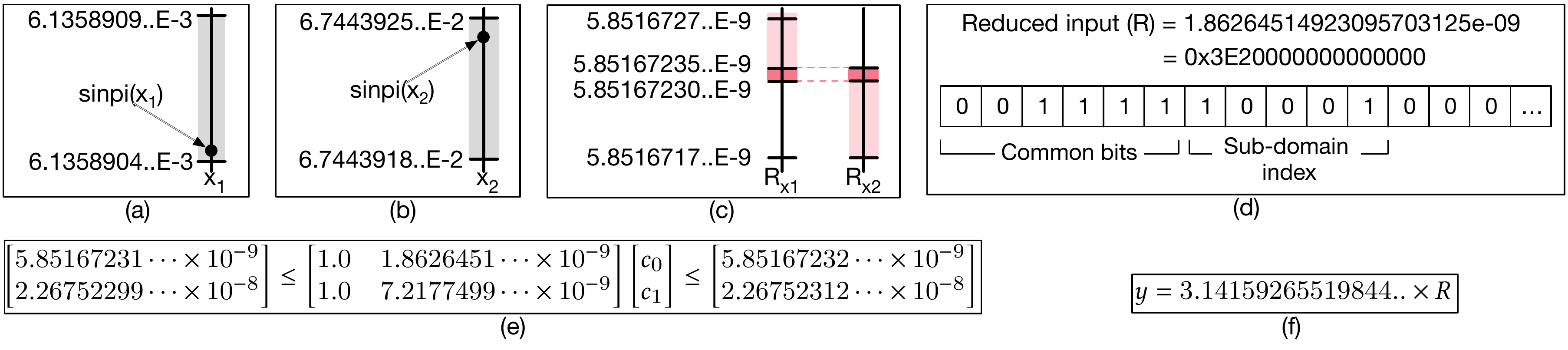}      
  \caption{\small (a) A 32-bit float input $x_1 = 1.95\dots \times
    10^{-3}$ and its correctly rounded result of $sinpi(x_1)$ (shown
    with a black circle). The rounding interval in double is colored
    gray. (b) Input $x_2 = 2.14\dots \times 10^{-2}$, its correctly
    rounded result for $sinpi(x_2)$, and the rounding interval. (c)
    The reduced intervals for $sinpi(R)$ corresponding to $x_1$ and
    $x_2$ so that both $x_1$ and $x_2$ produce correctly rounded
    results, respectively.  Both $x_1$ and $x_2$ map to the same
    reduced input $R$. The common interval for $R$ is highlighted with
    darker color.  (d) To approximate $sinpi(R)$, we create a
    piecewise polynomial with 32 sub-domains in total. We use the
    5-bits in the double representation of $R$ to identify the
    sub-domain for the piecewise polynomial. (e) Our LP formulation
    for generating a piecewise polynomial of degree 1 for the
    sub-domain with bit-pattern ($10001$) with two reduced inputs in
    the sample. (f) The resulting coefficients returned by the LP
    solver.}
  \label{fig:sinpiIntervalExample}
\end{figure*}

\subsection{Generating Piecewise Polynomials for $sinpi(x)$}
\label{sec:sinpi-overview}
To produce correctly rounded results for $sinpi(x)$, our approach
involves the following steps. First, we identify the correctly rounded
result and the rounding interval for each input in the entire domain.
Second, we identify the reduced rounding interval after range
reduction. Third, we split the reduced domain into sub-domains to
generate piecewise polynomials. Fourth, we perform counterexample
guided polynomial generation for each sub-domain. Finally, we validate
the generated piecewise polynomials for the entire input domain.

\vspace{4pt}
\noindent \textbf{Step 1: Identifying the correctly rounded result and
  the rounding interval.}  For each input $x$, we first identify the
correctly rounded result of $sinpi(x)$ using an oracle. Then, we
identify an interval of values $[l, h]$ in \texttt{double} where all
values in the interval rounds to the correctly rounded result. We call
this interval the rounding interval. If our polynomial approximation
produces a value in the rounding interval, the rounded result is the
correct result. Consider the inputs:

\begin{small}
\begin{align}
  x_1 &= 1.95312686264514923095703125 \times 10^{-3} \nonumber \\
  x_2 &= 2.148437686264514923095703125 \times 10^{-2} \nonumber
\end{align}
\end{small}

We show the correctly rounded result of $sinpi(x)$ for these inputs
with a black circle in Figure~\ref{fig:sinpiIntervalExample}(a) and
Figure~\ref{fig:sinpiIntervalExample}(b), respectively. It also shows
the rounding interval in gray.

\vspace{4pt}
\noindent \textbf{Step 2: Identifying the reduced interval for input
  R.}
Range reduction transforms input $x$ into $R$. To produce the result
for $sinpi(x)$, we need to compute both $sinpi(R)$ and $cospi(R)$
(\ie, multiple elementary functions). The result that we produce for
$sinpi(R)$ and $cospi(R)$ should allow us to produce the correct
result for $sinpi(x)$ (\ie, produce a value within the rounding
interval $[l, h]$ of input $x$).

To compute $sinpi(x)$, we will generate piecewise polynomials for
$sinpi(R)$ and $cospi(R)$. Two inputs $x_1$ and $x_2$
(Figure~\ref{fig:sinpiIntervalExample}(a) and
\ref{fig:sinpiIntervalExample}(b)) map to the same reduced input after
range reduction,

\begin{small}
\[
  R = 1.86264514923095703125 \times 10^{-9}
\]
\end{small}  

Now, we need to deduce an interval $[ls', hs']$ for the output of
$sinpi(R)$ and an interval $[lc', hc']$ for the output of $cospi(R)$
such that the result of output compensation produces a value within
the rounding interval for $x$.
We compute the correctly rounded value ($v$) of $sinpi(R)$ in double
using the oracle and set it as our initial guess for $[ls', hs']$
(\ie, $[v,v]$ a singleton). Similarly, we compute the interval $[lc',
  hc']$ for $cospi(R)$. Now, we need to check if these intervals are
sufficient to produce the correct output for the original input
$x$. Section~\ref{sec:reducedintervals} provides our detailed
algorithm. The key idea is to simultaneously lower the lower bound for
both $sinpi(R)$ and $cospi(R)$ and check if output compensation
produces the correct result for all inputs. Similarly, we deduce the
upper bound for both $sinpi(R)$ and $cospi(R)$.
The reduced interval for $sinpi(R)$ from the perspective of $x_1$ is
$[ls1', hs1']$. Similarly, the reduced interval for $sinpi(R)$ from
the perspective of $x_2$ is $[ls2', hs2']$.  These reduced intervals
for $sinpi(R)$ corresponding to $x_1$ and $x_2$ are shown in
Figure~\ref{fig:sinpiIntervalExample}(c).  They are not identical
because our approach considers the numerical error in both range
reduction and output compensation.

\vspace{4pt}
\noindent \textbf{Step 3: Splitting the reduced domain into
  sub-domains.} Now that we have reduced intervals for all reduced
inputs, the next task is to generate polynomials for $sinpi(R)$ and
$cospi(R)$. We illustrate this process with $sinpi(R)$. It is similar
for $cospi(R)$.  Even after range reduction, there are approximately
$110$ million unique reduced inputs for $R \in [0,
  \frac{1}{512}]$. Using our counterexample guided polynomial
generation strategy (Step 4), we attempt to generate a polynomial for
the entire reduced domain. If we cannot generate a polynomial or the
polynomial does not satisfy the performance constraints, then we split
the reduced input domain into smaller sub-domains to generate
piecewise polynomials. We iteratively split the domain into smaller
sub-domains until we can produce a polynomial that produces the
correct results for all inputs and satisfies the performance
criterion.

Let us say we want to generate 32~(\ie $2^5$) piecewise polynomials
for the domain $[0, \frac{1}{512}]$.  We use the bit-pattern of the
reduced input $R$ in double to identify the sub-domain. Although the
domain of R is $[0, \frac{1}{512}]$, the value of R in our reduced
inputs ranges between $[2^{-32}, 2^{-9}]$ along with $R = 0$.  There
is a large gap of values between the reduced input $0$ and
$2^{-32}$. This is because we have already handled special cases for
the original input.
Excluding the reduced input $R = 0$, all other reduced inputs in the
double representation have the left-most six bits identical. Hence, we
use 5-bits after the six left-most bits to identify the sub-domain for
the piecewise polynomial. Figure~\ref{fig:sinpiIntervalExample}(d)
shows the reduced input $R$, its double bit-pattern, and the 5-bits
used to identify the sub-domain.

\vspace{4pt}
\noindent \textbf {Step 4: Generating a polynomial for a sub-domain.}
The final step is to produce a polynomial that approximates $sinpi(R)$
for a particular sub-domain. This polynomial must produce a value
within the reduced interval $[ls', hs']$ for each reduced input $R$ in
the sub-domain. This requirement can be encoded as a linear constraint
for each reduced input,

\begin{small}
\[
  ls' \leq P(R) \leq hs'
\]
\end{small}
where $P(R)$ is a polynomial that approximates $sinpi(R)$.

We show the generation of the polynomial for sub-domain with
bit-pattern $10001$. First, we sample a portion of the reduced inputs
(\eg, 2 in Figure~\ref{fig:sinpiIntervalExample}(e)).  Second, we
encode the two reduced inputs and reduced intervals as linear
constraints to create a LP query (see
Figure~\ref{fig:sinpiIntervalExample}(e)). Third, we use a LP solver
to identify coefficients that satisfy the constraints. The generated
polynomial is shown in Figure~\ref{fig:sinpiIntervalExample}(f).
Fourth, we check if the generated polynomial produces a value within
the reduced interval for all inputs in the sub-domain.  In this case,
there are two reduced inputs where the generated polynomial does not
produce a value within the reduced interval. Fifth, we add both
counterexamples (\ie, reduced inputs) to the sample. Next, we create a
LP query using these four reduced inputs and intervals. Then, we check
if the generated polynomial satisfies all reduced inputs in the
sub-domain corresponding to bit-pattern $10001$.

After generating polynomials for all 32 sub-domains, we store the
coefficients of the piecewise polynomial in a table, which is indexed
by the bit-pattern of the reduced input that identifies the
sub-domain. The approximation for the elementary function $sinpi(x)$
is now ready.  To produce the result for input $x$, our library will
perform range reduction on $x$, identify the reduced input $R$,
identify the sub-domain based on the bit-pattern of $R$, evaluate the
piecewise polynomial using the coefficients from the table, perform
output compensation, and round the result to a 32-bit float to produce
the correctly rounded result.

\section{Generating Piecewise Polynomials}
Our goal is to generate polynomial approximations for elementary
functions $f(x)$ that produce the correctly rounded result for all
inputs $x$ in 32-bit target representations~$\mathbb{T}$. Similar to
our prior work on
\rlibm~\cite{lim:rlibm:popl:2021,Lim:rlibm:arxiv:2020}, we approximate
the correctly rounded result rather than the real value of $f(x)$. We
extend it in three main directions. First, we develop counterexample
guided polynomial generation with sampling to make this approach
feasible with 32-bit types. Second, we design techniques to generate
piecewise polynomials, which provide performance improvements. Third,
we develop modified range reduction techniques for a class of
elementary functions and develop methods to deduce rounding intervals
when range reduction involves multiple functions.

\textbf{Correctly rounded result.} For a given input $x$ and
elementary function $f$, the output of our approximation is the
correctly rounded result if it is equal to the value of $f(x)$
computed with real numbers and then rounded to the target
representation. We use $RN_{\mathbb{T}}(f(x))$ to denote the rounding
function that rounds $f(x)$ computed with real numbers to target
representation~$\mathbb{T}$. All internal computation such as range
reduction, polynomial evaluation, and output compensation is performed
in representation~$\mathbb{H}$ where $\mathbb{H}$ has higher precision
than $\mathbb{T}$. To attain good performance, $\mathbb{H}$ is a
representation that is supported in hardware (\eg, double).

\textbf{Our approach.} There are three main tasks in creating
polynomial approximations with our approach. First, we need to create
a range reduction function, which we denote as $RR_{\mathbb{H}}(x)$,
that reduces input $x$ to a reduced input $r$ in a smaller
domain. Once we have the result of the elementary function for the
reduced input $r$ (let's say $y'= f(r)$), we need to develop an output
compensation function, which we denote as $OC_{\mathbb{H}}(y', x)$, to
produce the result of $f(x)$ for input $x$. Second, we need to
generate polynomial approximations for each elementary
function~$f_i(r)$ in the reduced domain (\eg, there were two
elementary functions $sinpi$ and $cospi$ after range reduction in
Section~\ref{sec:overview}). We need to generate polynomials
$\Psi_{i}$ for each $f_i(r)$ in the reduced domain when there are
millions of reduced inputs in each reduced input domain.
We have to ensure that the polynomials generated for each $f_i(r)$ in
the reduced domain produce correctly rounded results for all inputs
after output compensation and polynomial evaluation is performed in
$\mathbb{H}$.
Third, we may have to split the reduced input domain to generate
piecewise polynomials for each $f_i(r)$ to create efficient
implementations.

\begin{algorithm}[t]
\small
\DontPrintSemicolon
\SetKwFunction{FCorrectPoly}{CorrectPolys}
\SetKwFunction{FCalcRndInterval}{RoundingInterval}
\SetKwFunction{FCalcReducedInterval}{ReducedIntervals}
\SetKwFunction{FGenApproxFunc}{GenApproxFunc}
\SetKwFunction{FCombineIntervals}{CommonIntervals}
\SetKwProg{Fn}{Function}{:}{}
\Fn{\FCorrectPoly{$f$, $X$, $RR_{\mathbb{H}}$, $OC_{\mathbb{H}}$,
    $d$}}{
  $Y \leftarrow \emptyset$\;
  \ForEach{$x \in X$} {
    $y \leftarrow RN_{\mathbb{T}}(f(x))$\;
    $[l, h] \leftarrow$ \FCalcRndInterval{$y$, $\mathbb{T}$, $\mathbb{H}$}\;
    $Y \leftarrow (x, [l, h])$
  }

  $\mathcal{L} \leftarrow $ \FCalcReducedInterval{$Y$, $RR_{\mathbb{H}}$, $OC_{\mathbb{H}}$}\;
  $Result \leftarrow \emptyset$\;
  
  \ForEach{$(f_i, \mathcal{L}_i) \in \mathcal{L}$} {
    \lIf{$\mathcal{L}_i \leftarrow \emptyset$} {
      \Return{$\emptyset$}
    }
    $\Psi_i \leftarrow $ \FGenApproxFunc{$\mathcal{L}_i$, $d$}\;
    $Result \leftarrow (f_i, \Psi_i) \cup Result$\;
  }
  \Return{$Result$}\;
}

\Fn{\FCalcRndInterval{$y$, $\mathbb{T}$, $\mathbb{H}$}}{
  $l \leftarrow min\{v \in \mathbb{H} \mid v \leq y \text{ and
    } RN_{\mathbb{T}}(v) = y\}$\;
  $h \leftarrow max\{v \in \mathbb{H} \mid v \geq y\text{ and
    } RN_{\mathbb{T}}(v) = y\}$  \;
  \Return{$[l, h]$}\;
}
  
\caption{\small \texttt{CorrectPolys} computes piecewise polynomials
  of degree $d$ for each elementary function $f_i$ used in output
  compensation, $OC_{\mathbb{H}}$, to generate a math library for
  elementary function $f$. It produces the correctly rounded result of
  $f(x)$ for each input $x \in X$.  \texttt{RoundingInterval} computes
  the rounding interval $[l, h] \subset \mathbb{H}$ of $y$ where all
  values in the interval rounds to $y$ in
  $\mathbb{T}$. \texttt{ReducedIntervals} is shown in
  Algorithm~\ref{alg:CalcReducedInterval} while \texttt{GenApproxFunc}
  is shown in Algorithm~\ref{alg:GenApproxFunc}.}
\label{alg:CorrectPoly}
\end{algorithm}

\textbf{High-level sketch.}  Algorithm~\ref{alg:CorrectPoly} provides
a high-level sketch of our approach. Given an elementary function
$f(x)$ and a list of inputs $X$, we compute the correctly rounded
result $y$ in our target representation $\mathbb{T}$ (line 4) and
compute the rounding interval of $y$ in $\mathbb{H}$ (lines 14-17). If
our approximation of $f(x)$ produces a value in the rounding interval,
then the result will round to $y$.
Next, we compute the reduced input $r$ using range reduction. The
range reduction may require us to compute multiple elementary
functions $f_i$ to produce the result for $x$.
Hence, we identify the range of values that each function $f_i$ should
produce such that the result when used with output compensation
produces a value in the rounding interval of $y$ (line 7). We call
this range of values for the reduced input $r$ as the reduced interval
(see Section~\ref{sec:reducedintervals}).
Finally, we approximate each elementary function $f_i(r)$ used in
output compensation with piecewise polynomials of degree $d$ (line 11)
with counterexample guided polynomial generation and by using an LP
solver. A single polynomial for each $f_i$ may not be ideal for
performance. To create efficient implementations, we iteratively split
the domain of the reduced input into multiple sub-domains~(see
Section~\ref{sec:approach:piecewise}). Even such sub-domains for the
reduced inputs can have millions of reduced inputs. Hence, we create a
sample of the reduced inputs, generate constraints to ensure that the
polynomial of degree $d$ produces a value in the reduced interval for
the reduced inputs in the sample, and query the LP solver to solve for
the coefficients. When the LP solver returns the coefficients, we
check whether the generated polynomial produces a value within the
reduced interval for all inputs in the sub-domain. We add any input
that violates the constraints to the sample and repeat this
process. We call this process as counterexample guided polynomial
generation. At the end of this process, our approach produces
piecewise polynomials for each $f_i(r)$, where the results of $f_i(r)$
when used with output compensation produces the correctly rounded
result for all inputs when rounded to $\mathbb{T}$.

\subsection{Computing Rounding Intervals}
Our approach approximates the correctly rounded result rather than the
real value. Hence, the first step is to identify the correctly rounded
result using an oracle and then identify all values in $\mathbb{H}$
that rounds to the correct result in $\mathbb{T}$. As $\mathbb{H}$ has
higher precision than $\mathbb{T}$, there is a range of values in
$\mathbb{H}$ that our approach can produce and still round to the
correctly rounded result in $\mathbb{T}$. We call this range the
rounding interval. Algorithm~\ref{alg:CorrectPoly} illustrates our
steps to compute the rounding interval for each input $x \in X$~(lines
14-17).
We compute the oracle correctly rounded result, $y=f(x)$, using the
MPFR math library with a large number of precision bits.  To compute
the rounding interval, we identify the smallest value $l \in
\mathbb{H}$ that rounds to $y$ when rounded to $\mathbb{T}$ and the
largest value $h \in \mathbb{H}$ that rounds to $y$ when rounded to
$\mathbb{T}$. This search procedure can be efficiently implemented
either using a binary search or by leveraging the properties of
$\mathbb{T}$ and $\mathbb{H}$. As long as our approach produces a
value in the rounding interval $[l, h]$ for input $x$, it will produce
the correctly rounded result.

\begin{algorithm}[t]
{\small
\DontPrintSemicolon
\SetKwFunction{FCorrectPoly}{CorrectPolynomials}
\SetKwFunction{FCalcRndInterval}{RoundingInterval}
\SetKwFunction{FGenApproxFunc}{GenApproxFunc}
\SetKwFunction{FCalcReducedInterval}{ReducedIntervals}
\SetKwFunction{FCombineIntervals}{CommonIntervals}
\SetKwComment{Comment}{$\triangleright$}{}
\SetKwFunction{FGetPrev}{GetPrev}
\SetKwFunction{FGetNext}{GetNext}
\SetKwProg{Fn}{Function}{:}{}
\Fn{\FCalcReducedInterval{$Y$, $RR_{\mathbb{H}}$, $OC_{\mathbb{H}}$}}{
  \lIf{$OC_{\mathbb{H}}$ is not a monotonic function} {
    \Return{$\emptyset$}
  }
  $F \leftarrow \{$list of functions used in $OC_{\mathbb{H}} \} $\;
  \lForEach{$f_i \in F$} {
    $\mathcal{L}_i \leftarrow \emptyset$
  }
  \ForEach{$(x, [l, h]) \in Y$} {
    $r \leftarrow RR_{\mathbb{H}}(x)$\;
    $V \leftarrow \{  RN_{\mathbb{H}}(f_i(r)) \mid f_i \in F  \}$ \;
    \lIf{$OC_{\mathbb{H}}(V, x) \notin [l, h]$} {
      \Return{$\emptyset$}
    }
    //Set initial reduced range for each $f_i(r)$\;
    $I' \leftarrow \{  [v, v] \mid v \in V  \}$ \;
    //Decrease the lower bounds $l_i'$ simulataneously\;
    \While{true} {
      $A \leftarrow \{$\FGetPrev{$l_i'$, $\mathbb{H}$} $\mid [l_i', h_i'] \in I'\}$\;
      \lIf{ $OC_{\mathbb{H}}(A, x) \notin [l, h]$ } {
        \textbf{break}
      }
      $I' \leftarrow \{  [$\FGetPrev{$I_i'$, $\mathbb{H}$}$, h_i'] \mid [l_i', h_i'] \in I'  \}$
    }
    //Increase the upper bounds $h_i'$ simulataneously\;
    \While{true} {
      $B \leftarrow \{$ \FGetNext{$h_i'$, $\mathbb{H}$} $\mid [l_i', h_i'] \in I'\}$\;
      \lIf{ $OC_{\mathbb{H}}(B, x) \notin [l, h]$ } {
        \textbf{break}
      }
      $I' \leftarrow \{  [I_i',$ \FGetNext{$h_i'$, $\mathbb{H}$}$] \mid [l_i', h_i'] \in I'  \}$
    }
    \ForEach{$[l_i', h_i'] \in I'$} {
      $\mathcal{L}_i \leftarrow \mathcal{L}_i \cup (r, [l_i', h_i'])$
    }
  }
  \Return{$\{  (f_i, \mathcal{L}_i) \mid f_i \in F  \}$}
}

\caption{\small \texttt{ReducedIntervals} computes the reduced
  interval $[l_i', h_i']$ and the reduced input $r$ corresponding to
  input $x$ for each function $f_i$ used with output compensation. If
  our polynomial approximation for $f_i$ produces a value in $[l_i',
    h_i']$, then we can generate the correctly rounded result for $x$.
  \texttt{ReducedIntervals} returns a list with $(r, [l_i', h_i'])$
  for each $f_i$. \texttt{GetPrev}(p, $\mathbb{H}$) returns the
  preceding value of $p$ in $\mathbb{H}$ and \texttt{GetNext}(p,
  $\mathbb{H}$) returns the succeeding value of $p$ in $\mathbb{H}$.
}
\label{alg:CalcReducedInterval}
}
\end{algorithm}

\subsection{Computing Reduced Rounding Intervals}
\label{sec:reducedintervals}
Range reduction is crucial for any technique that generates
approximations for elementary functions. It is particularly important
with our approach for 32-bit types because the condition number of the
LP problem increases drastically if the input domain has both
extremely large and small values. Further, large inputs can cause
overflows during the evaluation of a polynomial with a large degree in
$\mathbb{H}$.

After computing rounding intervals from
Algorithm~\ref{alg:CorrectPoly}, we have a list of constraints $(x,
[l, h])$ that our approximation for $f(x)$ needs to satisfy for each
input $x$ to produce the correctly rounded result. The range reduction
and subsequent output compensation can require us to approximate
multiple elementary functions $f_i$.  The next step is to identify
reduced inputs to $f_i$ and a range of values that $f_i$ should
produce such that the result of the output compensation produces a
value in $[l, h]$ for each $x$. The input to $f_i$ is the reduced
input and the range of values that $f_i$ should produce is the reduced
interval.

Algorithm~\ref{alg:CalcReducedInterval} shows the steps in deducing
the reduced interval. For each constraint $(x, [l, h])$, we can
identify the reduced input $r$ by performing range reduction on $x$
(line 6). However, computing the reduced interval is challenging.  We
present an algorithm to deduce reduced intervals when output
compensation ($OC_{\mathbb{H}}$) is monotonic (either increasing or
decreasing), which is the case with all range reductions that we
explore in the paper.

To compute the reduced interval, we identify all functions $f_i$ used
in $OC_{\mathbb{H}}$ (line 4). Then, we compute the correctly rounded
result $v_i$ for each $f_i(r)$ in $\mathbb{H}$ using an oracle (line
7).
If the result of output compensation using $v_i$'s does not produce a
value in the rounding interval for $x$, then either the range
reduction technique should be redesigned or the precision of
$\mathbb{H}$ should be increased.

Now, we have a candidate value (\ie, $v_i$) for each $f_i(r)$ to
produce the correctly rounded result of $x$. We have to deduce the
maximum amount of freedom available for each $f_i(r)$. We initially
set the reduced intervals $[l_i, h_i]$ for each $f_i$ to be $[v_i,
  v_i]$ (line 10).
Next, we identify if we can decrease the lower bound of the intervals
of $f_i(r)$. For a given reduced input $r$ of input $x$, we check if
using the preceding values of $l_i$ in $\mathbb{H}$ for all $f_i$'s
with output compensation produces a value in the rounding interval
$[l,h]$ of $x$.
If it does, then we widen the reduced interval by replacing each $l_i$
with the preceding value. We repeat the process until the result of
output compensation using the preceding values no longer produces a
value in $[l, h]$ (lines 12-15). This procedure to compute the lower
bound can be efficiently implemented by performing binary search
between $v_i$ and the minimum representable value.

Similarly, we identify if we can increase the upper bound of the
interval for each $f_i(r)$. For each upper bound $h_i$ of $f_i(r)$, we
identify the value that succeeds $h_i$ and check whether the result of
output compensation using the succeeding value produces a value in
$[l, h]$. If it does, then we widen the reduced interval by replacing
each $h_i$ with the succeeding value. We repeat the process until
output compensation produces a value outside the interval $[l,h]$ of
input $x$ (lines 17-20). The upper bound of the reduced interval can
be efficiently computed by performing binary search between $v_i$ and
the maximum representable value. Finally, we store the reduced
constraints $(r, [l_i', h_i'])$ for each function $f_i$ in a list
$\mathcal{L}_i$.

Each $\mathcal{L}_i$ corresponding to $f_i$ contains reduced intervals
$(r, [l_i', h_i'])$ for the reduced input $r$ to produce a correct result
for input $x$. As multiple inputs can map to the same reduced input
$r$, there can be multiple reduced constraints $(r, [l1_i', h1_i'])$
and $(r, [l2_i', h2_i'])$ for the same reduced input $r$ corresponding
to original inputs $x1$ and $x2$. The reduced intervals $[l1_i',
  h1_i']$ and $[l2_i', h2_i']$ are not exactly identical to account
for numerical errors in range reduction and output compensation.  Our
polynomial approximation for $f_i$ must satisfy the constraints $(r,
[l1_i', h1_i'])$ to produce the correctly rounded result for $x1$ and
$(r, [l1_i', h1_i'])$ to produce the correctly rounded result for
$x2$. Thus, we generate a single combined interval by computing the
common interval between them. If there is no common interval between
all reduced intervals corresponding to the same reduced input, then it
implies that there is no polynomial approximation for $f_i$ that
produces the correctly rounded results for all inputs $x$ in the
original domain. The library designer will have to redesign range
reduction in such cases.

\begin{algorithm}[t]
\small
\DontPrintSemicolon
\SetKwFunction{FCorrectPoly}{CorrectPoly}
\SetKwFunction{FCalcRndInterval}{CalcRndInterval}
\SetKwFunction{FGenApproxFunc}{GenApproxFunc}
\SetKwFunction{FCalcReducedInterval}{CalcReducedIntervals}
\SetKwFunction{FCombineIntervals}{CombineIntervals}
\SetKwFunction{FGetPrev}{GetPrev}
\SetKwFunction{FGetNext}{GetNext}
\SetKwFunction{FGenApproxFunc}{GenApproxFunc}
\SetKwFunction{FGenApproxHelper}{GenApproxHelper}
\SetKwFunction{FSplitDomain}{SplitDomain}
\SetKwFunction{FGenPiecewise}{GenPiecewise}
\SetKwFunction{FNumCommonBits}{NumCommonBits}
\SetKwFunction{FExtractBits}{Ext}
\SetKwFunction{FSampling}{GenPolynomial}
\SetKwProg{Fn}{Function}{:}{}
\Fn{\FGenApproxFunc{$\mathcal{L}$, $d$}}{
  $\mathcal{L}^{-} \leftarrow \{  (r, [l', h']) \in \mathcal{L} \mid r
  < 0\}$\;
  $\mathcal{L}^{+} \leftarrow \{  (r, [l', h']) \in \mathcal{L} \mid r
  \geq 0\}$\;
  $\Psi^{-} \leftarrow $ \FGenApproxHelper{$\mathcal{L}^{-}, d$}\;
  $\Psi^{+} \leftarrow $ \FGenApproxHelper{$\mathcal{L}^{+}, d$}\;
  \Return{$\{  \Psi^{-}, \Psi^{+}  \}$}
}

\Fn{\FGenApproxHelper{$\mathcal{L}$, $d$}}{
  $n \leftarrow 0$\;
  \While{true} {
    $\Delta = $ \FSplitDomain{$\mathcal{L}$, $n$}\;
    $(status, \Psi) = $ \FGenPiecewise{$\Delta$, $d$}\;
    \lIf{$status = true$} {
      \Return{$\Psi$}
    }
    $n \leftarrow n + 1$
  }
}

\Fn{\FGenPiecewise{$\Delta$, $d$}}{
  $\Psi \leftarrow \emptyset$\;
  \ForEach{$\Delta_j \in \Delta$} {
    $(status, \Psi_j) \leftarrow $ \FSampling{$\Delta_j$, $d$}\;
    \lIf{status = false} {
      \Return{$(false, \emptyset)$}
    }
    $\Psi \leftarrow \Psi \cup \Psi_j$\;
  }
  \Return{(true, $\Psi$)}
}
  
\caption{\small \texttt{GenApproxFunc} generates piecewise polynomials
  that produce a value in the reduced interval for all reduced inputs
  in $\mathcal{L}$.  It initially attempts to produce a single
  polynomial for the entire reduced input domain. If unsuccessful,
  then it splits the domain into multiple
  sub-domains. \texttt{SplitDomain} (not defined in this algorithm)
  splits the reduced input domain into sub-domains based on the
  bit-pattern of the reduced inputs in $\mathbb{H}$.
  \texttt{SplitDomain} returns $\Delta$, which includes a set of
  reduced constraints for each sub-domain $\Delta_j$.
  \texttt{GenPiecewise} generates a polynomial for each sub-domain,
  which is shown in Algorithm~\ref{alg:Sampling}.  }
\label{alg:GenApproxFunc}
\end{algorithm}

\subsection{Efficient Piecewise Polynomials }
\label{sec:approach:piecewise}
After the above steps, we have a list of reduced constraints $(r,
[l_i', h_i'])$ in $\mathcal{L}$ for each reduced input $r$ and for
each function $f_i$ that we need to approximate.  The next step in our
approach is to generate polynomials that approximate $f_i$ and satisfy
the constraints in $\mathcal{L}_i$.  Even after range reduction, there
can be hundreds of millions of reduced inputs.  The counterexample
guided polynomial generation algorithm, which we describe in
Section~\ref{sec:approach:sampling}, can likely generate a single
polynomial in many cases. However, it will also have a large degree and
may not be efficient. To generate high performance math libraries, we
propose the generation of piecewise polynomials. Effectively splitting
the domain into smaller domains for the generation of piecewise
polynomials is essential to improve performance. Hence, we group the
reduced input into sub-domains based on the bit-patterns of the
reduced input in $\mathbb{H}$.

Algorithm~\ref{alg:GenApproxFunc} describes our steps to generate
piecewise polynomials. Range reduction techniques for many elementary
functions can create both positive and negative reduced inputs (\eg,
$e^x$, $2^x$, $10^x$). The bit-patterns for positive and negative
reduced inputs in $\mathbb{H}$ will not have common bits at the
beginning (\eg, the explicit sign bit distinguishes positive and
negative values in double).  Hence, we separate the reduced inputs
(and their intervals) into two groups: $\mathcal{L}^-$ that contains
negative reduced inputs and $\mathcal{L}^+$ that contains non-negative
reduced inputs (lines 2-3). We create polynomial approximations for
each $\mathcal{L}^-$ and $\mathcal{L}^+$ (lines 4-5). This step also
allows us to subsequently group the reduced input into sub-domains in
an efficient manner.

If $\mathcal{L}$ contains only negative or positive reduced inputs, we
try to generate a single polynomial of degree $d$ that satisfies all
reduced constraints in $\mathcal{L}$ (line 11 and 17) using our
counterexample guided polynomial generation (see
Section~\ref{sec:approach:sampling}).  If it cannot generate a
polynomial of degree $d$ that satisfies all constraints, then we split
the reduced input domain in $\mathcal{L}$ into multiple sub-domains
(lines 9-13 in \texttt{GenApproxHelper}).
We iteratively split the domain of reduced inputs into $2^n$
sub-domains based on the bit-pattern of $r$ in $\mathbb{H}$ (\ie,
\texttt{SplitDomain} call in line 10).
To split the reduced input domain, we first identify the smallest
reduced input $R_{min}$ and the largest reduced input $R_{max}$. Then,
we compute the number of consecutive bits that are identical in the
bit-string representation of $R_{min}$ and $R_{max}$ in $\mathbb{H}$
starting from the most significant bit. We use the next $n$ bits to
identify the sub-domain for the piecewise polynomial.
Subsequently, we group the reduced inputs and reduced intervals based
on the bit-pattern of the reduced input into sub-domains ($\Delta$
returned by \texttt{SplitDomain}). We try to generate a polynomial of
degree $d$ that satisfies all reduced constraints in $\Delta_j$ for
all $\Delta_j$'s belonging to $f_i$ (lines 16-19).
Using bit-patterns of the reduced input in $\mathbb{H}$ allows us to
efficiently identify the sub-domain for the piecewise polynomial with
two bitwise operations (\texttt{and} and a \texttt{shift}). Once we
generate a polynomial for each sub-domain of every $f_i$, the
coefficients of the polynomial are stored in a table, which is indexed
using the bit-pattern of the reduced input for each $f_i$.

\begin{algorithm}[t]
\small
\DontPrintSemicolon
\SetKwFunction{FCorrectPoly}{CorrectPoly}
\SetKwFunction{FCalcRndInterval}{CalcRndInterval}
\SetKwFunction{FGenApproxFunc}{GenApproxFunc}
\SetKwFunction{FCalcReducedInterval}{CalcReducedIntervals}
\SetKwFunction{FCombineIntervals}{CombineIntervals}
\SetKwFunction{FGetPrev}{GetPrev}
\SetKwFunction{FGetNext}{GetNext}
\SetKwFunction{FGenApproxFunc}{GenApproxFunc}
\SetKwFunction{FGenApproxHelper}{GenApproxHelper}
\SetKwFunction{FSplitDomain}{SplitDomain}
\SetKwFunction{FGenPiecewise}{GenPiecewise}
\SetKwFunction{FNumCommonBits}{NumCommonBits}
\SetKwFunction{FExtractBits}{Ext}
\SetKwFunction{FSampling}{GenPolynomial}
\SetKwFunction{FSample}{Sample}
\SetKwFunction{FLPSolve}{GetCoeffsUsingLP}
\SetKwFunction{FCheckPoly}{Check}
\SetKwProg{Fn}{Function}{:}{}
\Fn{\FSampling{$\Delta_j$, $d$}}{
  $\mathcal{S} \leftarrow $ \FSample{$\Delta_j$}\;
  \While{true} {
    $\Psi_j \leftarrow$ \FLPSolve{$\mathcal{S}$, $d$}\;
    \lIf{$\Psi_j = \emptyset$} {
      \Return{$(false, \emptyset)$}
    }
    $(Done, \mathcal{S}) \leftarrow$ \FCheckPoly{$\Psi_j$,
      $\Delta_j$, $\mathcal{S}$}\;
    \lIf{$Done = true$} {
      \Return{$(true, \Psi_j)$}
    }
    \lIf{$|\mathcal{S}| > threshold$} {
      \Return{$(false, \emptyset)$}
    }
  }
}

\Fn{\FCheckPoly{$\Psi_j$, $\Delta_j$, $\mathcal{S}$}}{
  $Done \leftarrow true$\;
  \ForEach{$(r, [l', h']) \in \Delta_j$} {
    \If{not $l' \leq \Psi_j(r) \leq h'$} {
      $\mathcal{S} \leftarrow \{  (r, [l', h'])  \} \cup
      \mathcal{S}$\;
      $Done \leftarrow false$
    }
  }
  \Return{$(Done, \mathcal{S})$}
}
\caption{\small \texttt{GenPolynomial} attempts to find a polynomial
  of degree $d$ that satisfies the reduced input and interval
  constraints in $\Delta_j$ using our counterexample guided sampling
  approach. If it is infeasible to find a polynomial of degree $d$ or
  the size of the sample exceeds a threshold, then it returns (false,
  $\emptyset$). \texttt{GetCoeffsUsingLP} generates the coefficients
  of a polynomial that satisfies all constraints in $\mathcal{S}$
  using the LP solver. \texttt{Check} validates that the polynomial
  generated using the sample satisfies all reduced input and interval
  constraints. We add counterexamples (\ie, all inputs where the
  polynomial does not satisfy the constraints) to the sample and
  repeat the process.}
\label{alg:Sampling}
\end{algorithm}

\subsection{Counterexample Driven Polynomial Generation}
\label{sec:approach:sampling}
Once we have the reduced input and the reduced intervals, we structure
the problem of generating polynomials as a linear programming problem
similar to our prior work on
\rlibm~\cite{lim:rlibm:popl:2021,Lim:rlibm:arxiv:2020}.  Even after
range reduction and creation of sub-domains for the generation of
piecewise polynomials, we need to generate a polynomial approximation
when there are several million reduced inputs and reduced intervals in
the context of 32-bit types.  However, they are beyond the
capabilities of modern LP solvers, which can handle a few thousand
constraints. To address this issue, we propose counterexample guided
polynomial generation with sampling. The key insight is that we do not
need to add every reduced input and interval as a constraint in the LP
formulation as long as we identify and add the highly constrained
intervals.

Our counterexample guided polynomial generation strategy takes as
input the set of reduced constraints $(r, [l', h'])$ corresponding to
reduced inputs that belong to a particular sub-domain. The goal is to
generate a polynomial of degree $d$ that produces a value in the
reduced interval $[l', h']$ for each reduced input $r$. Each reduced
input $r$ and the corresponding interval $[l', h']$ specifies the
following linear constraint for a polynomial of degree $d$ that we
want to generate:

\begin{small}
\[
  l' \leq c_0 + c_1 r + c_2 r^2 + \dots + c_d r^d \leq h'
\]
\end{small}

The task of the polynomial generator is to find coefficients for the
polynomial.

To scale to 32-bit types, we sample a small fraction of the reduced
input and intervals. Algorithm~\ref{alg:Sampling} reports our
counterexample guided polynomial generation process.  It takes two
inputs: the degree of the polynomial and the set of reduced inputs and
intervals (\ie, $\Delta_j$) for generating a polynomial approximation
for an elementary function $f_i$ on reduced inputs for sub-domain~$j$.
We maintain the reduced inputs and their intervals in increasing
order.  Then we uniformly sample the reduced inputs based on the
distribution of reduced inputs. If there are a large number of reduced
inputs in a particular region of the sub-domain, then our method has
more samples from that region.  We also add highly constrained reduced
inputs and intervals (\ie, the correctly rounded result and the lower
bound/upper bound is less than $\epsilon$, which is set by the math
library designer) to the sample.

Then we express all constraints in the sample $(r, [l', h'])$ using a
single system of linear inequalities and solve for the coefficients
using an LP solver (line 4). If there are $n$ points in the sample,
the system of linear inequalities is of the following form:

\begin{small}
\[
  \begin{bmatrix}
    l_1' \\ l_2' \\ \vdots \\ l_n'
  \end{bmatrix}
  \leq
  \begin{bmatrix}
    1 & r_1 & \dots & r_1^d \\
    1 & r_2 & \dots & r_2^d \\
    \vdots & \vdots & \vdots & \vdots \\
    1 & r_n & \dots & r_n^d \\
  \end{bmatrix}
  \begin{bmatrix}
    c_0 \\ c_1 \\ \vdots \\ c_d
  \end{bmatrix}
  \leq
  \begin{bmatrix}
    h_1' \\ h_2' \\ \vdots \\ h_n'
  \end{bmatrix}
\]
\end{small}

There are two issues with the polynomial generated using the sampled
reduced inputs that we need to address. First, as the LP solver
returns coefficients as real numbers, the coefficients are rounded to
a value in $\mathbb{H}$.  As a result of rounding error, the result of
polynomial evaluation for a particular reduced input in the sample may
not lie within its rounding interval. Second, the polynomial generated
using the sample may not satisfy the constraints for the entire set of
reduced inputs and their corresponding intervals.

We address the real coefficients issue with a search-and-refine
procedure similar to \rlibm. When the LP solver returns real
coefficients and we round it to $\mathbb{H}$, we check whether
evaluating the polynomial satisfies constraints for every input in the
sample. If it does not, then we select the input and reduce its
rounding interval (either replace the lower bound with its succeeding
value or replace the upper bound with the preceding value). Then we
repeat the above process until it generates a polynomial that either
satisfies all constraints in the sample when evaluated in $\mathbb{H}$
or cannot find a polynomial of degree $d$. If we cannot find a
polynomial that satisfies all constraints in the sample, then we split
the entire reduced domain in $\mathcal{L}_i$ into even smaller
sub-domains and repeat this process.

If we successfully generate a polynomial $\Psi_j$ that satisfies all
constraints in the sample, then we check whether this polynomial
satisfies all constraints in $\Delta_j$ (line 10-15). If $\Psi_j$
satisfies all constraints, then we return the polynomial (line 7). If
there is any constraint not satisfied by $\Psi_j$ in the entire set of
reduced inputs, then we add that reduced input and its interval to the
sample (\ie, adding the counterexample in lines 12-13). We repeat the
process of generating the polynomial with the new sample. If the
number of constraints in the sample exceeds a threshold at any point,
then we determine that we cannot generate a polynomial for the
sub-domain $\Delta_j$.  Our function to generate the coefficients for
the polynomial (\ie, \texttt{GetCoeffsUsingLP}) using an LP solver
generates a polynomial of a lower degree (than input degree $d$) if it
is possible to do so.

\section{Experimental Evaluation}
\begin{table*}
  \small
  \caption{\small Generation of correctly rounded results for 32-bit
    floats with \tool, Intel's libm (float and double), glibc's libm
    (float and double), CR-LIBM, and MetaLibm (float and
    double). \cmark indicates that the library produces the correctly
    rounded result for all inputs. Otherwise, we use \xmark. For each
    \xmark, we show the number of inputs with wrong results. N/A
    indicates that the implementation is not available.}
  \begin{tabular}{| c | c | c | c | c | c | c | c | c |}
    \hline
    \begin{tabular}{@{}c@{}}\textbf{float} \\ functions\end{tabular}
 & \begin{tabular}{@{}c@{}}Using \\ \tool\end{tabular}
 & \begin{tabular}{@{}c@{}}Using \\ glibc float\end{tabular}
 & \begin{tabular}{@{}c@{}}Using \\ glibc double\end{tabular}
 & \begin{tabular}{@{}c@{}}Using \\ Intel float\end{tabular}
 & \begin{tabular}{@{}c@{}}Using \\ Intel double\end{tabular}
 & \begin{tabular}{@{}c@{}}Using \\ CR-LIBM\end{tabular}
 & \begin{tabular}{@{}c@{}}Using \\ MetaLibm float \end{tabular}
 & \begin{tabular}{@{}c@{}}Using \\ MetaLibm double \end{tabular}\\
\hline
\hline
$\mathbf{ln(x)}$ & \cmark & \xmark (4.2E5) & \xmark (5) & \xmark (1060) & \xmark (5)& \xmark (5) & N/A & N/A \\ \hline
$\mathbf{log2(x)}$ & \cmark & \xmark (3.1E5) & \cmark & \xmark (276) & \cmark& \cmark & N/A & N/A  \\ \hline
$\mathbf{log10(x)}$ & \cmark & \xmark (3.0E7) & \xmark (1) & \xmark (1.5E5) & \xmark (1)& \xmark (1) & N/A & N/A  \\ \hline
$\mathbf{exp(x)}$ & \cmark & \xmark (1.7E5) & \cmark & \xmark (2.5E5) & \cmark& \cmark  & \xmark (5.1E8) & \xmark (5.1E8)  \\ \hline
$\mathbf{exp2(x)}$ & \cmark & \xmark (1.7E5) & \xmark (2) & \xmark (7.2E5) & \xmark (2) & N/A & \xmark (6.5E7) & \xmark (1026)  \\ \hline
$\mathbf{exp10(x)}$ & \cmark & \xmark (1.7E5) & \cmark & \xmark (3.9E5) & \cmark & N/A & N/A & N/A  \\ \hline
$\mathbf{sinh(x)}$ & \cmark & \xmark (7.1E7) & \xmark (2) & \xmark (2.5E5) & \xmark (2) & \xmark (2) & N/A & N/A  \\  \hline
$\mathbf{cosh(x)}$ & \cmark & \xmark (1.8E7) & \cmark & \xmark (1.4E5) & \cmark & \cmark & \xmark (1.1E7) & \cmark  \\  \hline
$\mathbf{sinpi(x)}$ & \cmark & N/A & N/A & \xmark (3.4E5) & \cmark & \cmark & N/A & N/A  \\ \hline
$\mathbf{cospi(x)}$ & \cmark & N/A & N/A & \xmark (3.8E5) & \cmark & \cmark & N/A & N/A  \\ \hline
  \end{tabular}
  \label{tbl:floataccuracy}
\end{table*}

We provide details on our prototype, experimental methodology, and the
results of our experiments to check the correctness and performance of
the generated functions.

\subsection{Experimental Setup and Methodology}
\label{methodology}
\textbf{Prototype.}  The \tool prototype generates correctly rounded
elementary functions for 32-bit \texttt{floats} and \texttt{posit32},
which is a 32-bit posit type providing tapered precision (\ie, more
precision than float for values near
1)~\cite{Gustafson:online:2017:posit}. It contains ten correctly
rounded elementary functions for 32-bit floats and eight elementary
functions for the posit32 type.
To generate correctly rounded elementary functions with good
performance, the user can provide custom range reduction functions and
specify the degree or the structure of the polynomial (\ie, odd or
even).  \tool uses the MPFR library~\cite{Fousse:toms:2007:mpfr} with
up to 400 precision bits to compute the oracle for $f(x)$ and rounds
it to the target representation, which is good enough to compute the
oracle result for double~\cite{Lefevre:worstcase:arith:2001}.
\tool uses SoPlex~\cite{Gleixner:soplex:issac:2012}, an exact rational
LP solver, for generating coefficients for the polynomials with a five
minute time limit. We use a threshold of fifty thousand reduced
inputs and intervals in the sample for counterexample guided
polynomial generation. \tool's math library performs range reduction,
polynomial evaluation, and output compensation using double
precision. Polynomial evaluation uses the Horner's
method~\cite{borwein:polynomials:book:1995}. We designed novel
extensions to range reduction for many elementary functions, which is
inspired by table-based range reduction~\cite{Tang:exp:toms:1989,
  Tang:log:toms:1990, Tang:TableLookup:SCA:1991, Daramy:crlibm:doc}.
The appendix provides additional details about range reduction for
each elementary function. \tool is open source and publicly
available~\cite{rlibm-32}.
\textbf{Methodology.} We test the elementary functions in \tool on two
dimensions: (1) ability to generate correct results and (2)
performance in comparison to state-of-the-art libraries. We compare
\tool's functions with four libraries: Intel's libm, glibc's libm,
CR-LIBM~\cite{Daramy:crlibm:doc}, and
Metalibm~\cite{Olga:metalibm:icms:2014}. To use double precision
libraries, we convert the float input into double, use the double
function, and round the result back to float.  Among these libraries,
CR-LIBM has correctly rounded functions for double precision. However,
CR-LIBM does not produce correctly rounded results for 32-bit floats
due to double rounding.  There are no math libraries available for
posit32. All posit32 values can be exactly represented in double.
Hence, we compare our posit32 library with glibc and Intel's double
libm and CR-LIBM.

\textbf{Experimental setup.} We performed all our experiments on a
2.10GHz Intel Xeon Gold 6230R machine with 187GB of RAM running
Ubuntu~18.04. We disabled Intel turbo boost and hyper-threading to
minimize noise. We compiled \tool's math library at the \texttt{O3}
optimization level. We used Intel's libm from the oneAPI Toolkit and
glibc's libm from glibc-2.33.  We generated Metalibm implementations
with optimizations for \texttt{AVX2} extensions enabled.  Our test
harness that compares glibc's libm, CR-LIBM, and Metalibm with \tool
is built using the \texttt{gcc-10} compiler with \texttt{-O3 -static
  -frounding-math -fsignaling-nans} flags. To use Intel's libm, we
have to use the Intel compiler.  Hence, the test harness that compares
Intel libm with \tool is built using the \texttt{icc} compiler with
\texttt{-O3 -no-ftz -fp-model strict -static} to obtain as many
correct results as possible. Further, the size of the executable
generated by statically linking \tool is 2\% smaller on average when
compared to the executable generated with Intel's double libm.

\textbf{Measuring performance.} To compare performance, we measure the
number of cycles taken to compute the result for each input using
hardware performance counters. The total time taken is computed as the
sum of the time taken by all inputs (\ie, all $2^{32}$ inputs for a
32-bit representation). We ran the measurements for all inputs for
each function six times. Then, we compute the average time taken to
compute each elementary function.
As Intel's compiler performs vectorization by default at the
\texttt{O3} optimization level, our above setup does not measure
improvements due to vectorization.  Hence, we created another test
harness that creates an array of 1024 floats (\ie, $2^{10}$ inputs),
populates it with different inputs, and measures the number of cycles
taken to compute the results of $2^{10}$ inputs using hardware
performance counters. We repeat this experiment $2^{22}$ times to
compute the result and measure the total time taken for all $2^{32}$
inputs.

\subsection{Generation of Correctly Rounded Results}
Table~\ref{tbl:floataccuracy} reports the results of our experiments
to check the correctness of various elementary functions in \tool and
other mainstream libraries.  \tool produces the correctly rounded
results for all inputs for the ten elementary functions for 32-bit
floats. In contrast, elementary functions in glibc, Intel, and
MetaLibm's float library do not produce the correct result for all
inputs. Multiple functions in glibc and MetaLibm's float library
produce wrong results for several million inputs. Intel's libm also
produces wrong results with several thousand inputs with the float
version.
When we use double precision version of functions from glibc, Intel's
libm, and CR-LIBM, it does not produce the correct result for $ln(x)$,
$log10(x)$, $exp2(x)$, and $sinh(x)$. These cases occur when the real
value of $f(x)$ is extremely close to the rounding boundary of a
floating point value. Even with a smaller mini-max approximation error
in the double library compared to their float versions, these
libraries do not produce the correctly rounded result for all
inputs. CR-LIBM, which is a correctly rounded double library, produces
wrong results for float functions due to double rounding. We observed
that functions in MetaLibm do not produce correct results even when it
internally uses Sollya~\cite{Chevillard:sollya:icms:2010}, which can
be used to generate correctly rounded implementations.

\begin{table}
  \small
  \caption{\small Generation of correctly rounded results with posit32
    functions for all inputs by \tool, Intel and glibc's double
    libraries, and CR-LIBM. \cmark indicates that the library produces
    the correctly rounded result for all inputs and otherwise, we use
    \xmark.}
  \begin{tabular}{| c | c | c | c | c |} 
    \hline
    \begin{tabular}{@{}c@{}}\textbf{posit32} \\ functions\end{tabular}
& \begin{tabular}{@{}c@{}}Using \\ \tool\end{tabular}
& \begin{tabular}{@{}c@{}}Using \\ glibc double\end{tabular}
& \begin{tabular}{@{}c@{}}Using \\ Intel double\end{tabular}
& \begin{tabular}{@{}c@{}}Using \\ CR-LIBM\end{tabular}\\

    \hline
    \hline
    $\mathbf{ln(x)}$ & \cmark & \xmark (22) & \xmark (22) & \xmark (22) \\ \hline
    $\mathbf{log2(x)}$ & \cmark & \xmark (19) & \xmark (18) & \xmark (18)
    \\ \hline
    $\mathbf{log10(x)}$ & \cmark & \xmark (26) & \xmark (23) & \xmark (23)
    \\ \hline
    $\mathbf{Exp(x)}$ & \cmark & \xmark (4.4E8) & \xmark (4.4E8) &
                                                                   \xmark (4.4E8)
    \\ \hline
                
    $\mathbf{Exp2(x)}$ & \cmark & \xmark (4.0E8) & \xmark (4.0E8) & N/A \\ \hline
    $\mathbf{Exp10(x)}$ & \cmark & \xmark (5.2E8) & \xmark (5.2E8) & N/A \\ \hline
    $\mathbf{Sinh(x)}$ & \cmark & \xmark (4.4E8) & \xmark (4.4E8) & \xmark (4.4E8) \\ \hline
    $\mathbf{Cosh(x)}$ & \cmark & \xmark (4.4E8) &
                                                   \xmark (4.4E8) & \xmark (4.4E8) \\ \hline

  \end{tabular}
  \label{tbl:positaccuracy}
\end{table}

Table~\ref{tbl:positaccuracy} reports that \tool produces correctly
rounded results with all inputs for the eight posit32 functions.  All
posit32 values are representable in double precision but they cannot
be represented in 32-bit floats. Hence, we use CR-LIBM, Intel and
glibc's double library to compare with \tool. These libraries for
double precision do not produce correct results for all posit32
inputs. Unlike functions for 32-bit floats, they produce wrong results
for several million inputs especially for exponential and hyperbolic
functions. One of the key reasons for wrong results is the absence of
overflows to $\infty$ and underflows to 0 with the posit32
type. Instead, extremely large values are rounded to the largest
representable value. Similarly, extremely small values are rounded to
the smallest non-zero representable value in the posit32 type.
  
\begin{table}
  \small
  \caption{\small Details about the generated polynomials. For each
    elementary function, time taken to generate the polynomials in
    minutes, the size of the piecewise polynomial for approximating
    $f_i(r)$, the maximum degree of the polynomial, and the number of
    terms in the polynomial.}
    \begin{tabular}{| c | c | c | c | c | c |} 
      \hline
	$f(x)$
      & \begin{tabular}{@{}c@{}}Gen. Time \\ (Minutes)\end{tabular}
      & \begin{tabular}{@{}c@{}}Reduced \\ Inputs\end{tabular}
      & \begin{tabular}{@{}c@{}}\# of Poly- \\ nomials \end{tabular}
      & \begin{tabular}{@{}c@{}}Deg-\\ree\end{tabular} 
      & \begin{tabular}{@{}c@{}}\# of \\ Terms\end{tabular} \\
      \hline
      \hline
      \multicolumn{6}{| c |}{\textbf{float functions}} \\ \hline
      $ln(x)$  	& $218$ 		& 7.2E6 & $2^{10}$ & 3 & 3	\\ \hline
      $log2(x)$  	& $251$ 	& 7.2E6 & $2^{8}$ & 3 & 3	\\ \hline
      $log10(x)$  	& $429$ 	& 7.2E6 & $2^{8}$ & 3 & 3	\\ \hline
      $exp(x)$  	& $117$ 			& 5.2E8 & 
      	\begin{tabular}{@{}c@{}}$2^{7}$ \\ $2^{7}$\end{tabular} & 
	\begin{tabular}{@{}c@{}} 4 \\ 4\end{tabular} & 
	\begin{tabular}{@{}c@{}} 5 \\ 5\end{tabular}\\ \hline
      $exp2(x)$  	& $86$ 			& 3.0E8 & 
      	\begin{tabular}{@{}c@{}}$2^{4}$ \\ $2^{3}$\end{tabular}  & 
	\begin{tabular}{@{}c@{}} 4 \\ 4\end{tabular} & 
	\begin{tabular}{@{}c@{}} 5 \\ 5\end{tabular}\\ \hline
      $exp10(x)$ 	& $169$ 		& 5.2E8 & 
      	\begin{tabular}{@{}c@{}}$2^{6}$ \\ $2^{7}$\end{tabular}  & 
	\begin{tabular}{@{}c@{}} 4 \\ 3\end{tabular} & 
	\begin{tabular}{@{}c@{}} 5 \\ 4\end{tabular}\\ \hline
      $sinh(x)$  	& $28$ 		& 1.5E8 & $2^6$ & 5 & 3	\\  
      $cosh(x)$  	& $24$ 		& 1.5E8 & $2^6$ & 4 & 3	\\  \hline
      $sinpi(x)$  	& $30$ 		& 1.2E8 & 1 & 5 & 3	\\ 
      $cospi(x)$  	& $19$ 		& 1.2E8 & 1 & 4 & 3	\\ \hline

      \hline
      \multicolumn{6}{| c |}{\textbf{posit32 functions}} \\ \hline
      $ln(x)$  	& 264 & 1.1E8 & $2^{11}$ & 4 & 4  \\ \hline
      $log2(x)$  	& 288 & 1.1E8 & $2^{8}$ & 4 & 4  \\ \hline
      $log10(x)$  	& 685 & 1.1E8 & $2^{12}$ & 3 & 3  \\ \hline
      $exp(x)$  	& 1089 & 3.5E9 &
      \begin{tabular}{@{}c@{}}$2^{12}$ \\ $2^{12}$\end{tabular}  &
      \begin{tabular}{@{}c@{}}$3$ \\ $3$\end{tabular} &
      \begin{tabular}{@{}c@{}}$4$ \\ $4$\end{tabular}   \\ \hline
      $exp2(x)$  	& 814 & 7.9E8 &
      \begin{tabular}{@{}c@{}}$2^{10}$ \\ $2^{12}$\end{tabular}  &
      \begin{tabular}{@{}c@{}}$3$ \\ $3$\end{tabular} &
      \begin{tabular}{@{}c@{}}$4$ \\ $4$\end{tabular}   \\ \hline
      $exp10(x)$	& 1528 & 3.4E9 &
      \begin{tabular}{@{}c@{}}$2^{13}$ \\ $2^{13}$\end{tabular}  &
      \begin{tabular}{@{}c@{}}$3$ \\ $3$\end{tabular} &
      \begin{tabular}{@{}c@{}}$4$ \\ $4$\end{tabular}   \\ \hline
      $sinh(x)$  	& 461 & 1.6E9 &
      \begin{tabular}{@{}c@{}}$2^{14}$ \\ $2^{14}$\end{tabular}  &
      \begin{tabular}{@{}c@{}}$5$ \\ $4$\end{tabular} &
      \begin{tabular}{@{}c@{}}$3$ \\ $3$\end{tabular}   \\ \hline
      $cosh(x)$  	& 528 & 1.7E9 &
      \begin{tabular}{@{}c@{}}$2^{14}$ \\ $2^{12}$\end{tabular}  &
      \begin{tabular}{@{}c@{}}$3$ \\ $6$\end{tabular} &
      \begin{tabular}{@{}c@{}}$2$ \\ $4$\end{tabular}   \\ \hline
    \end{tabular}
\label{tbl:statistics}
\end{table}

\textbf{Piecewise polynomials generated by \tool.}
Table~\ref{tbl:statistics} provides details on the piecewise
polynomials generated by \tool.
Our goal is to get the best possible performance within a given
storage budget for piecewise polynomials (\ie, number of sub-domains
when we split the range of reduced inputs). Hence, we used the \tool
to generate piecewise polynomials such that the degree of each
polynomial was less than or equal to 8 and the number of sub-domains
was less than or equal to $2^{14}$. The output compensation for
$sinh(x)$, $cosh(x)$, $sinpi(x)$, and $cospi(x)$ involves two
elementary functions. We generate two piecewise polynomials for each
of those elementary functions.
There are both positive and negative reduced inputs for $exp(x)$,
$exp2(x)$, and $exp10(x)$. Hence, we created two piecewise
polynomials: one for the negative reduced inputs and another for
positive reduced inputs.
Notably, we were able to generate a single polynomial of degree 5 and
4 that satisfies all reduced constraints for $sinpi(r)$ and
$cospi(r)$, respectively. Both $sinpi(r)$ and $cospi(r)$ have close to
120 million reduced inputs. Our counterexample driven polynomial
generation with sampling was instrumental in creating this efficient
polynomial.

\textbf{Time taken to generate \tool functions.}
Table~\ref{tbl:statistics} also reports the time taken to generate the
32-bit float and the posit32 functions in \tool. It ranges from 19
minutes for $cospi(x)$ for the float type to approximately 25 hours
for $exp10(x)$ for the posit32 type.  Majority of the total time total
time is spent in computing the oracle result and the rounding interval
using the MPFR library (\ie, 86\% of total time for 32-bit floats and
55\% of total time for the posit32 type). In contrast, counterexample
guided polynomial generation takes 14\% and 45\% of the total time for
32-bit floats and the posit32 type, respectively.  We noticed that it
takes significantly longer to generate posit32 functions. There are
fewer special cases, which requires longer oracle
computation. Further, \tool generates larger piecewise polynomials for
posit32 functions to account for higher precision than a 32-bit float
and saturating behavior with extremal values.

\begin{figure*}
  \small
  \begin{subfigure}[b]{0.99\columnwidth}
    \caption{Speedup of \tool's float functions over glibc libm}
    \includegraphics[width=\linewidth]{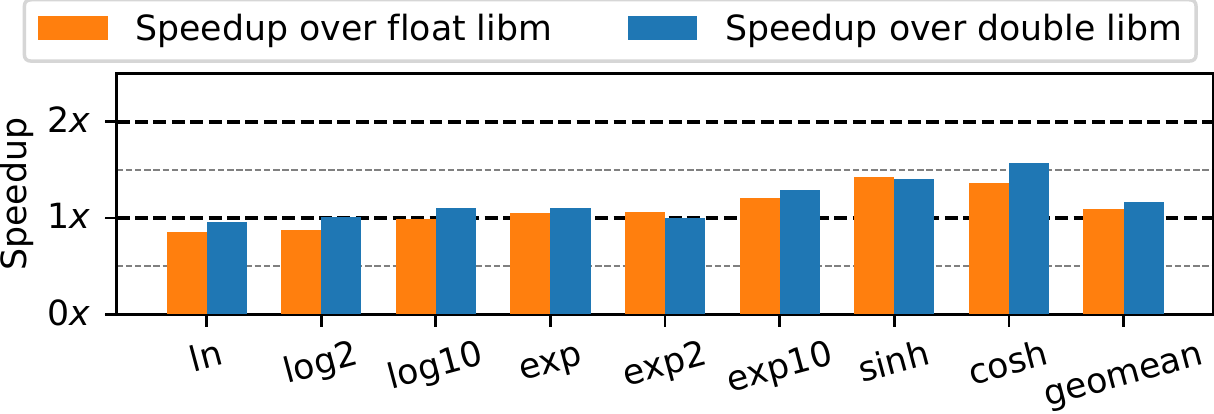}
  \end{subfigure}
  \hfill
  \begin{subfigure}[b]{0.99\columnwidth}
    \caption{Speedup of \tool's float functions over Intel libm}
    \includegraphics[width=\linewidth]{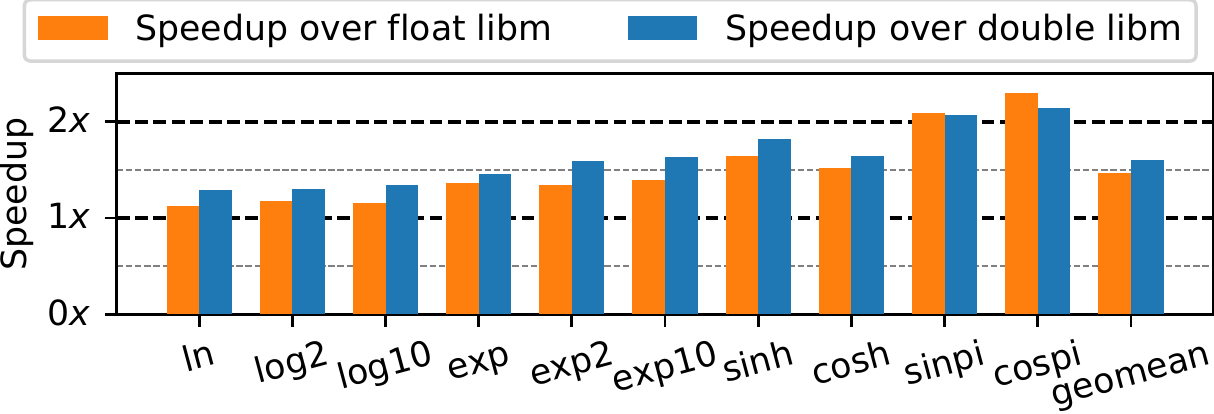}
  \end{subfigure}
  
  \vspace{1em}
  
  \begin{subfigure}[b]{0.99\columnwidth}
    \caption{Speedup of \tool's float functions over CR-LIBM}
    \includegraphics[width=\linewidth]{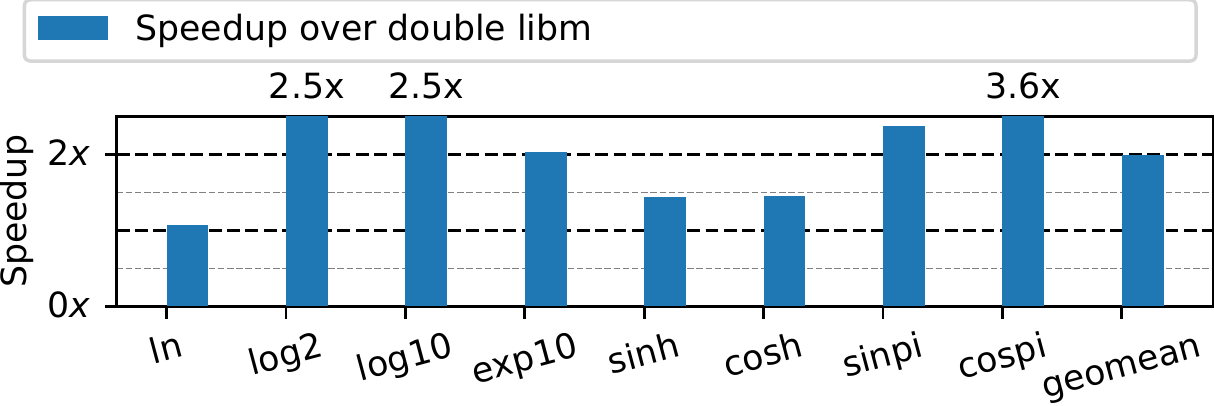}
  \end{subfigure}
  \hfill
  \begin{subfigure}[b]{0.99\columnwidth}
    \caption{Speedup of \tool's float functions over MetaLibm}
    \includegraphics[width=\linewidth]{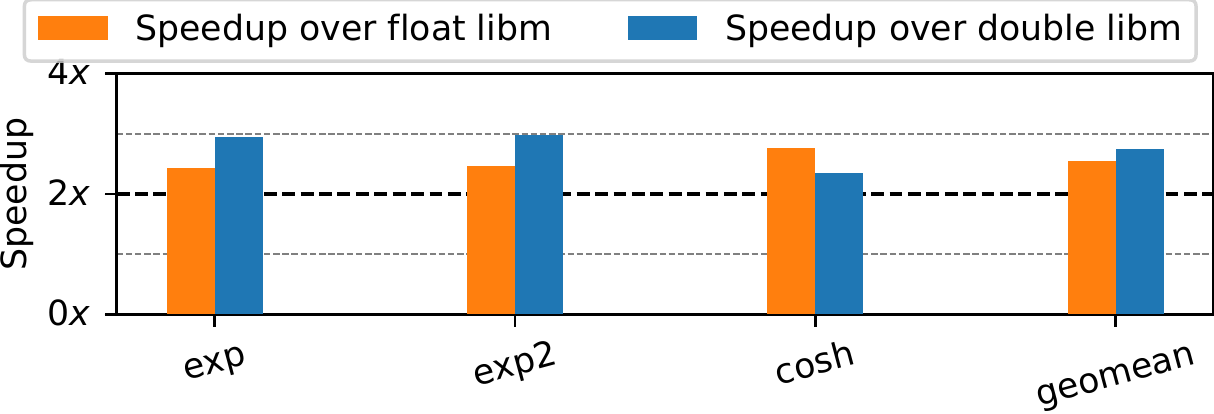}
  \end{subfigure}
  \caption{\small (a) Speedup of \tool's float functions compared to
    glibc's float functions (left) and glibc's double functions
    (right). (b) Speedup of \tool's functions compared to Intel's
    float functions (left) and Intel's double functions (right). (c)
    Speedup of \tool's functions compared to CR-LIBM functions. (d)
    Speedup of \tool's functions compared to MetaLibm's float
    functions (left) and double functions (right) built with AVX2
    optimizations.}
  \label{fig:floatSpeedup}
\end{figure*}

\subsection{Performance Evaluation of \tool}

\textbf{Performance of float functions.}
Figure~\ref{fig:floatSpeedup}(a) presents the speedup of \tool's float
functions over glibc's float functions (left bar in each cluster) and
double functions (right bar in each cluster). On average, \tool's
float functions have $1.1\times$ speedup over glibc's float libm and
$1.2\times$ speedup over glibc's double
libm. Figure~\ref{fig:floatSpeedup}(b) reports the speedup of \tool's
float functions over Intel's float libm and double libm. \tool's float
functions have an average of $1.5\times$ speedup over Intel's float
functions and $1.6\times$ speedup over Intel's double functions.
Figure~\ref{fig:floatSpeedup}(c) reports that \tool's functions are on
average $2\times$ faster than CR-LIBM functions.
Figure~\ref{fig:floatSpeedup}(d) reports the speedup of \tool's
functions over MetaLibm's float and double functions. \tool's
functions are on average $2.5\times$ and $2.7\times$ faster than
MetaLibm's float and double functions, respectively.
\tool's functions are faster than all the corresponding functions in
Intel libm, CR-LIBM, and MetaLibm.  \tool's functions are faster than
glibc's functions except for $ln(x)$, $log_2(x)$, and $log_{10}(x)$
for float and $ln(x)$ for double.
However, glibc's libm produces a large number of wrong results for
them.
\tool's functions are not only faster but also produce correctly
rounded results for all inputs.

\begin{figure*}
\centering
\captionsetup{justification=centering}
  \small
  \begin{subfigure}[b]{0.34\textwidth}
    \caption{Speedup of \tool's posit32 functions over glibc libm}
    \includegraphics[width=\linewidth]{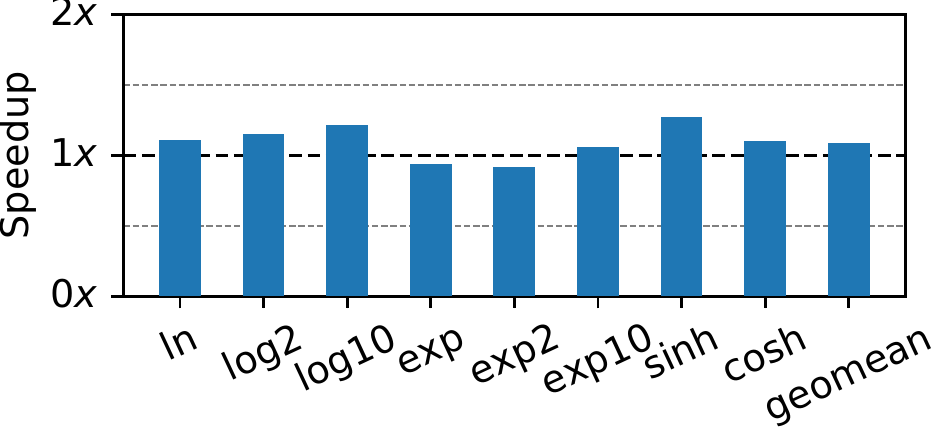}
  \end{subfigure}
  \hfill
  \begin{subfigure}[b]{0.34\textwidth}
    \caption{Speedup of \tool's posit32 functions over Intel libm}
    \includegraphics[width=\linewidth]{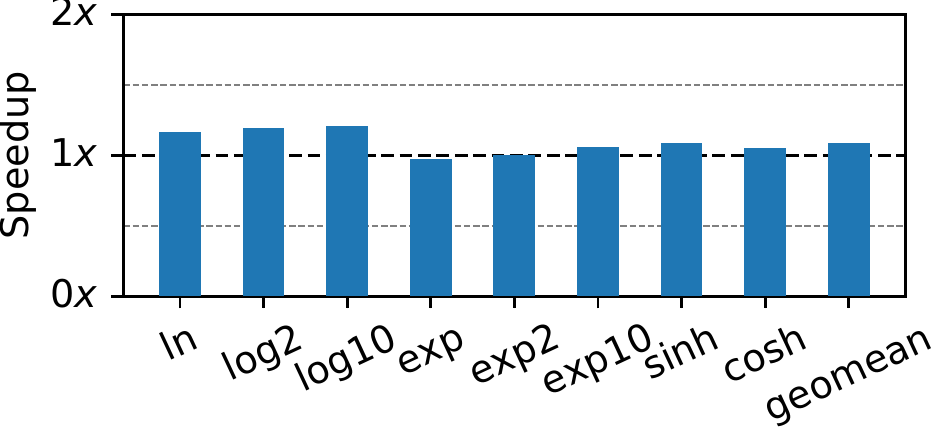}
  \end{subfigure}
  \hfill
  \begin{subfigure}[b]{0.27\textwidth}
    \caption{Speedup of \tool's posit32 functions over CR-LIBM}
    \includegraphics[width=\linewidth]{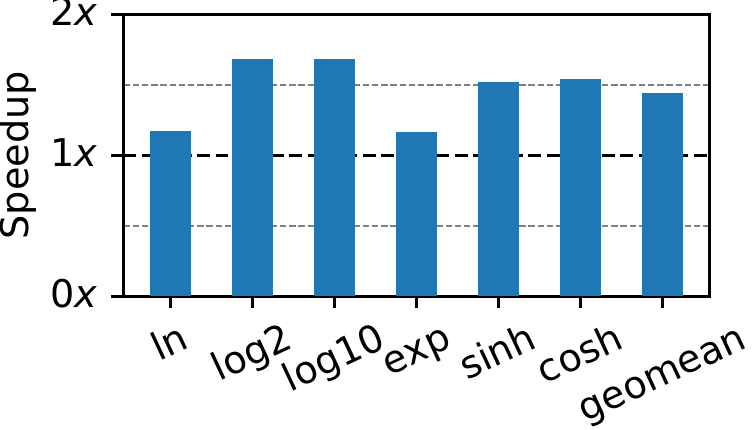}
  \end{subfigure}
  \caption{\small (a) Speedup of \tool's posit32 functions compared to
    glibc's double functions. (b) Speedup of \tool's posit32 functions
    compared to Intel's double functions. (c) Speedup of \tool's
    posit32 functions compared to CR-LIBM functions.}
  \label{fig:positSpeedup}
\end{figure*}

\textbf{Performance of posit32 functions.}  The graphs in
Figure~\ref{fig:positSpeedup}(a), Figure~\ref{fig:positSpeedup}(b),
and Figure~\ref{fig:positSpeedup}(c) report the speedup of \tool's
posit32 functions when compared to math libraries created by
re-purposing glibc's, Intel's, and CR-LIBM's double functions,
respectively. On average, \tool's posit32 functions are $1.1\times$,
$1.1\times$, and $1.4\times$ faster than glibc's libm, Intel's libm,
and CR-LIBM, respectively. All three re-purposed math libraries
produce wrong results for some inputs. \tool provides the first
correctly rounded functions for the posit32 type.  

\textbf{Vectorization.} Intel compiler uses vector instructions to
improve performance by default.  In our experiments with vectorization
using an array of 1024 inputs (see Section~\ref{methodology}), \tool
is on average 10\% and 5\% slower than Intel's float libm and double
libm, respectively. However, Intel's compiler produces wrong results
for several million inputs (without \texttt{-no-ftz -fp-model strict}
flags). In contrast, \tool's functions are almost as fast as
vectorized code while producing correct results for all inputs.

\begin{figure}
    \small
    \includegraphics[width=\linewidth]{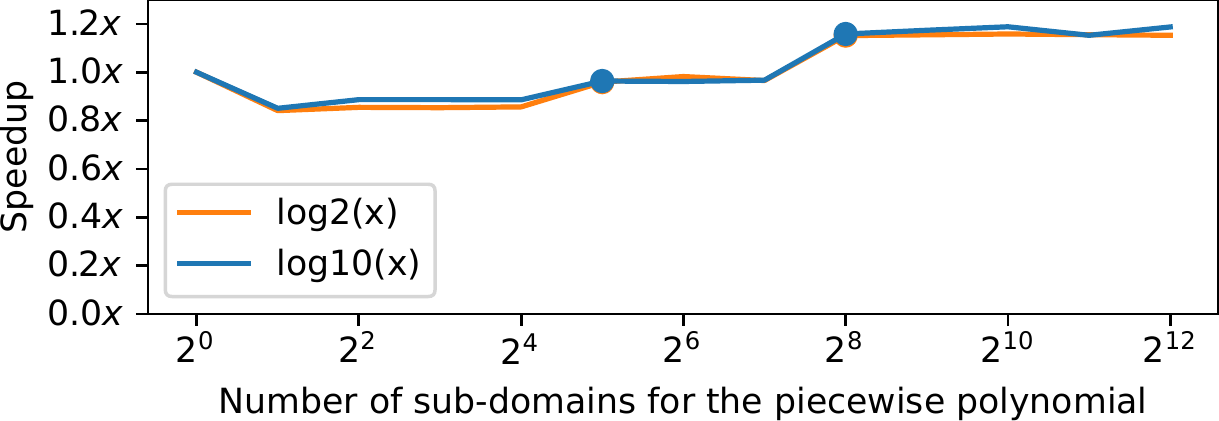}
    \caption{\small Performance speedup of $log2(x)$ and $log10(x)$
      with an increase in the number of sub-domains when compared to a
      single polynomial generated by \tool. All these polynomials
      produce the correctly rounded result for all inputs. A circle
      represents a decrease in the degree of the piecewise
      polynomial.}
    \label{fig:piecewiseSpeedup}
\end{figure}

\textbf{Performance impact of piecewise polynomials.} To analyze the
performance benefits due to piecewise polynomials, we identified
elementary functions for which we could generate a single polynomial
that produces correctly rounded results for all inputs ($log2(x)$,
$log10(x)$, $sinpi$, and $cospi$) . We measured the change in
performance with an increase in the number of sub-domains ranging from
$2^0$ (\ie, a single polynomial) to $2^{12}$.
Figure~\ref{fig:piecewiseSpeedup} reports the performance of $log2(x)$
and $log_{10}(x)$ with an increase in the number of sub-domains when
compared to the performance of a single polynomial. We validated that
all these polynomials produce the correct result for all
inputs. Figure~\ref{fig:piecewiseSpeedup} does not report $sinpi$ and
$cospi$ because the single polynomial has the best performance.
Initially, there is a small decrease in performance by moving
from a single polynomial to a piecewise polynomial because the degree
of the piecewise polynomial does not decrease significantly to subsume
the overhead of table lookup. On increasing the number of sub-domains,
we observed almost $1.2\times$ speedup with piecewise polynomials
having $2^8$ sub-domains. It requires 6\texttt{KB} for storing
coefficients of piecewise polynomials.

\section{Case Study with $cospi(x)$ for Float}
\label{sec:casestudy}
We describe the case study with $cospi(x) = cos (\pi x)$ to illustrate
the importance of carefully designing range reduction to avoid
cancellation errors in output compensation. The elementary function
$cospi(x)$ is defined for $x \in (-\infty, \infty)$.

\textbf{Special cases.}  There are three kinds of special cases:

\begin{small}
\[
  cospi(x) =
  \begin{cases}
    1.0 & \text{if } |x| < 7.771 \times 10^{-5}\\
    (-1)^{(|x| \:mod \: 2)} \times 1.0 & \text{if } |x| \geq 2^{23}\\
    NaN & \text{if } x = NaN \text{ or } x = \pm \infty \\
  \end{cases}
\]
\end{small}

All float values $\geq 2^{23}$ are integers. Hence, $cospi(x)
= 1.0$ for even integers and $cospi(x) = -1.0$ for odd integer inputs.

\textbf{Range reduction of $cospi(x)$.}  After excluding special
cases, there are more than 600 million float inputs that need
to be approximated. Similar to range reduction for $sinpi(x)$
(Section~\ref{overview:rangereduction}), we use periodicity and
trigonometric identities of $cospi(x)$ to reduce inputs to a smaller
domain. We transform input $x$ into $x = 2.0 \times I + J$ where $I$
is an integer and $J \in [0, 2)$. Due to periodicity, $cospi(x) =
  cospi(J)$. Next, we decompose $J$ into $J = K + L$ where $K$ is the
  integral part of $J$ ($K \in \{0, 1\}$) and $L \in [0, 1)$ is the
    fractional part. Then, $cospi(J)$ can be computed with,

\begin{small}
\[
  cospi(J) = (-1)^{K} cospi(L)
\]
\end{small}

To further reduce the range of $L$, we use the fact that $cospi(x)$
between $[0.5, 1)$ is a mirror image of $cospi(x)$ between $[0, 0.5)$
\textbf{with the opposite sign}. We decompose $L$ into $M$ and $L'$
where

\begin{small}
  \[
    M =
    \begin{cases}
      0 & \text{if } \leq 0.5 \\
      1 & \text{if } L > 0.5
    \end{cases}
    \quad\quad
    L' =
    \begin{cases}
      L & \text{if } L \leq 0.5 \\
      1.0 - L & \text{if } L > 0.5
    \end{cases} 
  \]
\end{small}

We have $cospi(L) = (-1)^M cospi(L')$. After reducing the input $x$ to
$L' \in [0, 0.5]$, there are around 107 million inputs. Thus, we
further reduce $L'$ to a value in $[0, \frac{1}{512}]$. We split
$L' = \frac{N}{512} + Q$ where $N$ is an integer in the set
$\{0, 1, 2, \dots, 255\}$ and $Q$ is a fractional value in
$[0, \frac{1}{512}]$.
One possible method to compute $cospi(L')$ is to use the trigonometric
identity $cospi(a + b) = cospi(a) cospi(b) - sinpi(a) sinpi(b)$,

\begin{small}
\[
cospi\left(\frac{N}{512} + Q\right) = cospi\left(\frac{N}{512}\right) cospi(Q) -
sinpi\left(\frac{N}{512}\right) sinpi(Q)
\]
\end{small}

The above formula is not monotonic and can have cancellation errors if
$N \neq 0$ (if $N = 0$, then $cospi(L') = cospi(Q)$).

\textbf{Creating monotonic output compensation.}  If $N \neq 0$, we
transform $N$ and $Q$ to $N'$ and $R$ such that $L' = \frac{N'}{512} -
R$ to create a monotonic output compensation function:

\begin{small}
\[
  N' =
  \begin{cases}
    0 & \text{if } N = 0 \\
    N + 1 & otherwise
  \end{cases}
  \quad\quad
  R =
  \begin{cases}
    Q & \text{if } N = 0 \\
    \frac{1}{512} - Q & otherwise
  \end{cases}
\]
\end{small}

Then, we can compute $cospi(L')= cospi(\frac{N'}{512} - R)$ using the
trigonometric identity
$cospi(a-b) = cospi(a) cospi(b) + sinpi(a) sinpi(b)$ as follows,

\begin{small}
  \begin{align}
    &cospi(L') = \nonumber \\ 
    &\begin{cases}
      cospi(R) & \text{if } N = 0 \\
      cospi\left(\frac{N'}{512}\right) cospi(R) +
      sinpi\left(\frac{N'}{512}\right) sinpi(R) & \text{if } N \neq 0
    \end{cases}\nonumber
\end{align}
\end{small}

This output compensation is monotonic and does not experience
cancellation error. The values of $N'$ ranges from $0$ to $256$ and $R
\in [0, \frac{1}{512}]$.  The computation $\frac{1}{512} - Q$ can be
computed exactly with float or double type for all
values of $Q$ that corresponds to $N \neq 0$. There are approximately
40 million values of $R$. We precompute the values for
$cospi\left(\frac{N'}{512}\right)$ and
$sinpi\left(\frac{N'}{512}\right)$ in lookup tables (\ie, 514 values
in total). We create polynomial approximations for $sinpi(R)$ and
$cospi(R)$ for the reduced input domain $R \in [0,
  \frac{1}{512}]$. Using \tool, we were able to generate a single
$5^{th}$ degree odd polynomial for $sinpi(R)$ and a single $4^{th}$
degree even polynomial for $cospi(R)$.  Finally, we can compute the
result for $cospi(x)$ with the output compensation function,

\begin{small}
\[
  cospi(x) =
  \begin{cases}
    S \times cospi(R) & \text{if } N = 0 \\
    S \times
  \left(cpn \times cospi(R) +
    spn \times sinpi(R)\right) & \text{if } N
  \neq 0
  \end{cases}
\]
\end{small}

where $S = (-1)^K \times (-1)^M$, $cpn = cospi(\frac{N'}{512})$, and
$spn = sinpi(\frac{N'}{512})$.
These polynomials combined with the output compensation functions
produce correctly rounded results for all inputs for $sinpi(x)$ and
$cospi(x)$.

\section{Related Work}
Multiple decades of seminal work has
advanced the state-of-the-art on creating approximations for FP
representations~\cite{Jeannerod:sqrt:tc:2011,
  Bui:exp:ccece:1999,
  Abraham:fastcorrect:toms:1991,
  Fousse:toms:2007:mpfr,Muller:elemfunc:book:2005,Trefethen:chebyshev:book:2012,
  Remes:algorithm:1934, Caro:polynomial:tcas:2017}. Further, seminal
research on range reduction has made such approximation
feasible~\cite{Tang:log:toms:1990,Tang:TableLookup:SCA:1991,Tang:exp:toms:1989,762822,
  Cody:book:1980, Boldo:reduction:toc:2009}.  Simultaneously, there
are verification efforts to prove bounds for math
libraries~\cite{harrison:hollight:tphols:2009,
  Harrison:expproof:amst:1997, Harrison:verifywithHOL:tphol:1997,
  Sawada:verify:acl:2002, Lee:verify:popl:2018}, identify numerical
errors with expressions that can be used in the implementation of math
libraries~\cite{Chowdhary:positdebug:2020:pldi,
  Benz:pldi:2012:dynamic, Fu:weakdistance:pldi:2019, Goubalt:2001:sas,
  Sanchez:pldi:2018:herbgrind}, and repair individual outputs of math
libraries~\cite{Panchekha:herbie:pldi:2015,
  Xin:repairmlib:popl:2019,Daming:fpe:popl:2020}.

\textbf{Correctly rounded libraries.} Numerous groups have developed
correctly rounded elementary functions~\cite{Jeannerod:sqrt:tc:2011,
  Bui:exp:ccece:1999}. Some correctly rounded libraries for FP are IBM
LibUltim~\cite{
  Abraham:fastcorrect:toms:1991}, Sun Microsystem's
LibMCR,
CR-LIBM~\cite{Daramy:crlibm:doc}, MPFR math
library~\cite{Fousse:toms:2007:mpfr}, and
RLIBM~\cite{lim:rlibm:popl:2021,Lim:rlibm:arxiv:2020}.
CR-LIBM is a
correctly rounded double library developed using
Sollya~\cite{Chevillard:sollya:icms:2010}, which generates mini-max
polynomials to approximate elementary
functions~\cite{Brisebarre:maceffi:toms:2006,
  Brisebarre:epl:arith:2007}. Sollya uses the modified Remez
algorithm~\cite{Remes:algorithm:1934} using lattice basis reduction
and also computes the error bound of the
polynomial~\cite{Chevillard:infnorm:qsic:2007, Chevillard:ub:tcs:2011,
  Guillaume:Gappa:online:2019}. Metalibm~\cite{Olga:metalibm:icms:2014,
  Brunie:metalibm:ca:2015} builds on Sollya and generates efficient
mini-max polynomials with user-defined error bounds. It also uses
domain splitting and hardware specific
optimizations~\cite{Olga:split:scan:2015}. Compared to mini-max
approaches, our work approximates the correctly rounded result of
$f(x)$ and generates polynomials that already account for numerical
error in range reduction and output compensation. Hence, it generates
efficient and correctly rounded results for all inputs.

This paper extends our prior work on
\rlibm~\cite{lim:rlibm:popl:2021,Lim:rlibm:arxiv:2020} and John
Gustafson's Minefield method~\cite{Gustafson:unum:2020:online}, which
advocate approximating the correctly rounded value rather than real
value of an elementary function. Our prior work on \rlibm also frames
the problem of generating polynomials as an LP problem.  We have used
\rlibm to create correctly rounded functions for 16-bit types:
bfloat16 and posit16. This paper extends \rlibm to handle 32-bit types
with systematic counterexample guided polynomial generation,
generation of piecewise polynomials to improve performance, and new
techniques to deduce rounding intervals when range reduction involves
multiple elementary functions.

\textbf{Posit libraries.}
SoftPosit-Math~\cite{Leong:online:2019:positmath} and \rlibm libraries
provide correctly rounded math functions for 16-bit posits. In our
prior work, we have produced approximations for a set of trigonometric
functions using the CORDIC method for
posit32~\cite{Lim:cordic:cf:2020}. However, it does not produce
correct results for all inputs.  In this paper, we develop the first
set of elementary functions that produce correctly rounded results for
all inputs for 32-bit posits.

\section{Conclusion and Future Directions}
Mainstream math libraries have been designed and improved by numerous
researchers spanning multiple decades. Yet, they fail to generate
correct results for all inputs.  This paper advocates approximating
the correctly rounded value instead of the real value similar to our
prior work on \rlibm. It extends \rlibm to scale to 32-bit
representations: (a) counterexample guided polynomial generation with
an LP solver to handle billions of inputs, (b) generation of
constraints to account for multiple elementary functions in range
reduction, and (c) generation of piecewise polynomials. The resulting
functions produce correct results for all inputs and are also faster
than existing libraries for 32-bit floats and posits.

Going forward, we plan to generate approximations for all commonly
used elementary functions with 32-bit types, which we believe can be
accomplished with our approach. However, it may require us to develop
novel extensions to range reduction. Further, it may be necessary to
perform range reduction in higher precision for some trigonometric
functions such as sine and cosine that use $\pi$. Beyond 32-bit types,
we also plan to extend this approach to double precision. Our approach
can generate a polynomial that produces the correctly rounded result
for the sampled points in the double type. Validating the correctness
of the result produced by a polynomial generated using our approach
for all inputs in the double type is an open research problem.  Our
long-term goal is to enable the standards of existing and new
representations to mandate correctly rounded results.

\begin{acks}                            
  We thank our shepherd Rahul Sharma and the PLDI reviewers for their
  feedback.
  We thank John Gustafson for his inputs on the Minefield
  method and the posit representation.
  This material is based upon
  work supported in part by the \grantsponsor{GS100000001}{National Science
    Foundation}{http://dx.doi.org/10.13039/100000001} under Grant
  No.~\grantnum{GS100000001}{1908798} and Grant
  No.~\grantnum{GS100000001}{1917897}.

\end{acks}
\balance
\bibliography{reference}
\clearpage
\appendix
\section{Details on Range Reduction used in \tool}
\label{apx:General}
In this section, we describe the special cases, the range reduction,
and the output compensation functions that we used to create the math
library functions in \tool.  A given family of elementary functions
(\eg, $ln(x)$, $log_2(x)$, and $log_{10}(x)$) uses similar approaches
for range reduction.  Hence, we first describe the range reduction
technique we used for each family of elementary functions. In the
subsequent sections, we describe the special cases and specific
details on range reduction and output compensation that we used for
float functions (Appendix~\ref{apx:float}) and posit32 functions
(Appendix~\ref{apx:posit}).

\subsection{Logarithm Functions $log_a(x)$}
\label{apx:log}
We use a table-based approach~\cite{Tang:log:toms:1990} to perform
range reduction for $log_a(x)$ functions. The range reduction is
generally applicable to any value of $a$. To reduce the input $x$ to
the reduced input in a smaller domain, we first transform input $x$
into $x = m \times 2^n$ where $n$ is an integer representing the
exponent of $x$ and $m \in [0, 1)$ is the significand. Then,
$log_a(x)$ can be computed with,
\[
  log_a(x) = log_a(m \times 2^n) = log_a(m) + nlog_a(2)
\]

Next, we further reduce the range of $m$ by transforming $m$ into
$m = F + f$ where $F = 1 + \frac{J}{128}$, $J$ is a value in the set
$\{0, 1, 2, \dots, 127\}$, and $f \in [0, \frac{1}{128})$ is the
remaining value. Intuitively, F is the value represented by the first
8 bits of the significand $m$ and $f$ is the value represented by the
rest of the bits. Then, $log_a(m)$ can be computed with,
\begin{small}
  \begin{align}
    log_a(m) &= log_a\left(F + f\right) = log_a\left(F \: \left(1 +
    \frac{f}{F}\right)\right) \nonumber \\
    &= log_a(F) + log_a\left(1
    + \frac{f}{F}\right) \nonumber
  \end{align}
\end{small}
We can denote $r = \frac{f}{F}$. Then,
$log_a(m)$ can be computed as,
\begin{small}
  \[
    log_a(m) = log_a(F) + log_a\left(1 + r\right)
  \]
\end{small}

The reduced input $r$ is in the range of $[0, \frac{1}{128})$. The
computation $r = \frac{f}{F}$ can be efficiently performed by
computing $f \times \frac{1}{F}$ if the value $\frac{1}{F}$ can be
computed ahead of time. Thus, we pre-compute the values for $log_a(F)$
(128 values for each $log_a$) and $\frac{1}{F}$ (128 values in total)
in lookup tables. We approximate $log_a(1 + r)$ for the reduced input
domain $r \in [0, \frac{1}{128})$. Finally, we can compute the result
of $log_a(x)$ with the output compensation function,
\begin{small}
  \[
    log_a(x) = log_a(1 + r) + log_a(F) + nlog_a(2)
  \]
\end{small}

\subsection{Exponential Functions $a^x$}
\label{apx:exp}
We also use a table-based approach~\cite{Tang:exp:toms:1989} to
perform range reduction for $a^x$ functions. The range reduction is
applicable for different values of $a$. To reduce the input $x$ we
first transform $x$ into $x = nlog_a(2) + \frac{j}{64}log_a(2) + r$
where $n$ is an integer, $j$ is a value in a set
$\{0, 1, 2, \dots, 63\}$ and $|r| \leq
\frac{log_a(2)}{64}$. Intuitively, $n$ represents the integral part of
the value $\frac{x}{log_a(2)}$. The value $\frac{j}{64}$ represents
the first 6 fractional bits of $\frac{x}{log_a(2)}$. Lastly, $r$
represents the remaining value,
$r = x - nlog_a(2) - \frac{j}{64}log_a(2)$.

The scaling by $log_a(2)$ allows us to create efficient output
compensation formula. The value of $a^x$ can be computed using the
property $a^{x + y} = a^xa^y$ and $a^{xlog_a(y)} = y^x$ as,
\begin{small}
\begin{align}
  a^x = a^{nlog_a(2) + \frac{j}{64}log_a(2) + r} &= a^{nlog_a(2)} \times
                                             a^{\frac{j}{64}log_a(2)} \times a^{r} \nonumber \\
  &= 2^n \times 2^{\frac{j}{64}} \times a^{r} \nonumber
\end{align}
\end{small}
Multiplication by $2^n$ can be computed efficiently using bit-wise
operations. We pre-compute and store the value of $2^{\frac{j}{64}}$ in
a table (\ie 64 values in total) and we approximate $a^{r}$ for the input
domain of $r \in [-\frac{log_a(2)}{64}, \frac{log_a(2)}{64}]$.

\subsection{Sinpi(x)}
\label{apx:sinpi}
The range reduction technique for $sinpi(x)$ first leverages the
periodicity of $sinpi(x)$ to reduce the input and then use a
table-based range reduction~\cite{Tang:TableLookup:SCA:1991} to
further reduce the input. First, we transform input $x$ into
$x = 2i + j$ where $i$ is an integer and $j \in [0, 2)$. Then,
$sinpi(x) = sinpi(j)$. Next, we decompose $j$ into $j = k + l$ where
$k \in \{0, 1\}$ is the integral part of $j$ and $L \in [0, 1)$ is the
fractional part. Then, $sinpi(j)$ can be computed with,
\begin{small}
  \[
    sinpi(j) = (-1)^k \times sinpi(l)
  \]
\end{small}
Third, we use the fact that $sinpi$ between $[0.5, 1)$ is a mirror
image of values between $[0, 0.5]$ and decompose $l$ into,
\begin{small}
  \[
    l' = \begin{cases}
      l & \text{if } l \leq 0.5 \\
      1.0 - l & \text{if } l > 0.5
      \end{cases}
  \]
\end{small}
Then, $sinpi(l)$ can be computed with $sinpi(l) = sinpi(l')$.

We further reduce $l'$ to a value between $[0, \frac{1}{512}]$ using
table-based range reduction. We split $l'$ into $l' = \frac{n}{512} +
r$ where $n$ is an integer in the set $\{0, 1, \dots, 255\}$ and $r$
is a real value in $[0, \frac{1}{512}]$. The value of $sinpi(l')$ can
be computed using the trigonometric identity $sinpi(a + b) = sinpi(a)
cospi(b) + cospi(a) + sinpi(b)$:
\begin{small}
  \[
    sinpi(l') = sinpi\left(  \frac{n}{512}  \right)cospi(r) + cospi\left(  \frac{n}{512}  \right)sinpi(r)
  \]
\end{small}

We pre-compute and store the values of $sinpi\left( \frac{n}{512} \right)$ and
$cospi\left( \frac{n}{512} \right)$ in lookup tables (\ie 512
values). We approximate $sinpi(r)$ and $cospi(r)$ for the reduced
input domain $r \in [0, \frac{1}{512}]$. Finally, we can approximate
$sinpi(x)$ using the output compensation function,
\begin{small}
  \[
    sinpi(x) = (-1)^k \times \left(  sinpi\left(  \frac{n}{512}  \right)cospi(r) + cospi\left(  \frac{n}{512}  \right)sinpi(r)  \right)
  \]
\end{small}

\subsection{Cospi(x)}
\label{apx:cospi}
Similar to $sinpi(x)$, we leverage the periodicity of $cospi(x)$ and
table-based range reduction~\cite{Tang:TableLookup:SCA:1991} to
reduce the input. Additionally, we apply some modifications to create
monotonic output compensation function. First, we transform the input
$x$ into $x = 2i + j$ where $i$ is an integer and $j \in [0, 2)$. Due
to periodicity, $cospi(x) = cospi(j)$. Second, we decompose $j$ into
$j = k + l$ where $k \in \{0, 1\}$ is the integral part of j and $l
\in [0, 1)$ is the fractional part. Then, $cospi(j)$ can be computed
with,
\begin{small}
  \[
    cospi(j) = (-1)^k \times cospi(l)
  \]
\end{small}

Third, we reduce $l$ using the fact that $cospi(x)$ between $[0.5, 1)$
is a mirror image of $cospi(x)$ between $[0, 0.5]$ with the opposite
sign. We decompose $l$ into $m$ and $l'$ where:
\begin{small}
  \[
    m =
    \begin{cases}
      0 & \text{if } l \leq 0.5 \\
      1 & \text{if } l > 0.5
    \end{cases}
    \quad\quad
    l' =
    \begin{cases}
      l & \text{if } l \leq 0.5 \\
      1.0 - l & \text{if } l > 0.5
    \end{cases}
  \]
\end{small}
Then, $cospi(l)$ can be computed with $cospi(l) = (-1)^m cospi(l')$.

We further reduce $l'$ to a value between $[0, \frac{1}{512}]$ using
table-based range reduction. We transform $l'$ into $r$ and $n$ using,
\begin{small}
  \[
    l' =
    \begin{cases}
      r & \text{if } l' < \frac{1}{512} \\
      \frac{n}{512} - r & otherwise
    \end{cases}
  \]
\end{small}
where $n$ is an integer value in the set $\{0, 1, 2, \dots, 256\}$ and
$r$ is a fractional value in $[0, \frac{1}{512}]$. Then, $cospi(l')$
can be computed with the trigonometric identity $cospi(a - b) =
cospi(a) copsi(b) + sinpi(a) sinpi(b)$,
\begin{small}
  \[
    cospi(l') =
    \begin{cases}
      cospi(r) & \text{if } l' < \frac{1}{512} \\
      cpn \times cospi(r) + spn \times sinpi(r) & otherwise
    \end{cases}
  \]
\end{small}
where $cpn = cospi\left( \frac{n}{512} \right)$ and
$spn = sinpi\left( \frac{n}{512} \right)$. This formula is monotonic
for all inputs $x$. We pre-compute and store the values of
$sinpi\left( \frac{n}{512} \right)$ and
$cospi\left( \frac{n}{512} \right)$ in lookup tables (\ie 514
values). We approximate $sinpi(r)$ and $cospi(r)$ for the reduced
input domain $r \in [0, \frac{1}{512}]$.

Finally, we compute the result of $cospi(x)$ with the output
compensation formula,
\begin{small}
  \[
    cospi(x) =
    \begin{cases}
      s \times cospi(r) & \text{if } l' < \frac{1}{512} \\
      s \times (cpn \times cospi(r) + spn \times sinpi(r)) & otherwise
    \end{cases}
  \]
\end{small}
where $s = (-1)^k \times (-1)^m$.

\subsection{Sinh(x)}
\label{apx:sinh}
We use the range reduction technique similar to the table-based range
reduction technique used in CR\_LIBM. First, the $sinh(x)$ function
has a property, $sinh(-x) = -sinh(x)$. Thus, the result of $sinh(x)$
for $x < 0$ can be derived by computing $-1 \times sinh(|x|)$.

Next, we decompose $|x|$ into three parts:
\begin{small}
\[
  |x| = k {ln(2)} + \frac{j}{64} ln(2) + r
\]
\end{small}
where both $k$ and $j$ are integers, $k \geq 0$, $0 \leq j < 64$, and
$r$ is a value in $[0, \frac{ln(2)}{64})$. As we will show later,
scaling by $ln(2)$ allows us to compute $sinh(k{ln(2)})$ and
$cosh(k{ln(2)})$ efficiently with minimal amount of error. If we
denote $K = k {ln(2)}$ and $J = \frac{j}{64} ln(2)$, then $sinh(|x|)$
can be computed using the hyperbolic identities,
$sinh(a + b) = sinh(a) cosh(b) + cosh(a) sinh(b)$ and
$cosh(a + b) = cosh(a) cosh(b) + sinh(a) sinh(b)$:
\begin{small}
\begin{align}
  sinh(|x|) &= SH \times cosh(r) + CH \times sinh(r)  \nonumber \\
  SH &= sinh(K) \times cosh(J) + cosh(K) \times sinh(J) \nonumber\\ 
  CH &= cosh(K)  \times cosh(J) + sinh(K)  \times sinh(J) \nonumber
\end{align}
\end{small}

We store the values of $sinh(K)$, $cosh(K)$, $sinh(J)$, and $cosh(J)$
in lookup tables. Storing $sinh(J)$ and $cosh(J)$ requires a total of
128 values. The number of values to store for $sinh(K)$ and $cosh(K)$
depends on the target representation. In the case of float,
the values of $k$ does not exceed $129$ because $|x| < 130 ln(2)$ for
all non-special-case inputs. Thus, we store $260$ values for
$sinh(K)$ and $cosh(K)$.

Alternatively, we can choose to not store the values of $sinh(K)$ and
$cosh(K)$ by computing $sinh(K)$ and $cosh(K)$ manually using the
following properties,
\begin{small}
\[
  sinh(x) = \frac{e^x - e^{-x}}{2} \quad \quad cosh(x) = \frac{e^x + e^{-x}}{2}
\]
\end{small}
Then, $sinh(K)$ and $cosh(K)$ can be computed as follows,
\begin{small}
\begin{align}
  sinh(K) = sinh(k ln(2)) = 2^{k - 1} - 2^{-k - 1} \nonumber \\ 
  cosh(K) = cosh(k ln(2)) = 2^{k - 1} + 2^{-k - 1} \nonumber
\end{align}
\end{small}
Although $sinh(K)$ and $cosh(K)$ cannot be exactly represented by
double type for all $K$, the correctly rounded value of $sinh(K)$ and $cosh(K)$
can be computed efficiently using integer arithmetic, floating point
subtraction, and addition. We plan to incorporate this logic in our
implementation of $sinh(x)$ to reduced the size of the lookup table.

We approximate $sinh(r)$ and $cosh(r)$ for the reduced inputs
$r \in [0, \frac{ln(2)}{64})$. Finally, $sinh(x)$ can be computed with
the output compensation function,
\begin{small}
\[
  sinh(x) = s \times \left(SH \times cosh(r) + CH \times sinh(r)\right)
\]
\end{small}
where $s$ is the sign of $x$.

\subsection{Cosh(x)}
\label{apx:cosh}
The range reduction technique for $cosh(x)$ uses a similar technique
as $sinh(x)$ (described in Appendix~\ref{apx:sinh}) to reduce the
input $x$. Then it uses the hyperbolic identities of $cosh(x)$ to
perform output compensation. First, the $cosh(x)$ function
has a property, $cosh(-x) = cosh(x)$. Thus, the result of $cosh(x)$
for $x < 0$ can be derived by computing $cosh(|x|)$.

Next, we decompose $|x|$ into three parts similar to how we decompose
the input $|x|$ for $sinh(x)$:
\begin{small}
\[
  |x| = k {ln(2)} + \frac{j}{64} ln(2) + r
\]
\end{small}
Both $k$ and $j$ are integers, $k \geq 0$, $0 \leq j < 64$, and $r$ is
a real number value in $[0, \frac{ln(2)}{64})$. If we denote
$K = k {ln(2)}$ and $J = \frac{j}{64} ln(2)$, then $cosh(|x|)$ can be
computed using the hyperbolic identities,
$sinh(a + b) = sinh(a) cosh(b) + cosh(a) sinh(b)$ and
$cosh(a + b) = cosh(a) cosh(b) + sinh(a) sinh(b)$:
\begin{small}
\begin{align}
  cosh(x) &= SH \times sinh(R) + CH \times cosh(R)  \nonumber \\
  SH &= sinh(K) \times cosh(J) + cosh(K) \times sinh(J) \nonumber\\ 
  CH &= cosh(K)  \times cosh(J) + sinh(K)  \times sinh(J) \nonumber
\end{align}
\end{small}

We store the values of $sinh(K)$, $cosh(K)$, $sinh(J)$, and $cosh(J)$
in lookup tables (\ie 388 values). We approximate $sinh(r)$ and
$cosh(r)$ for the reduced inputs $r \in [0,
\frac{ln(2)}{64})$. Finally, $cosh(x)$ can be computed with the output
compensation function,
\begin{small}
\[
  cosh(x) = SH \times sinh(r) + CH \times cosh(r)
\]
\end{small}

\section{Details on 32-bit Float Functions}
\label{apx:float}
In this section, we explain the 32-bit float functions in
\tool. We describe the special cases and any specific details on the
range reduction technique used for each function.

\subsection{$ln(x)$}
The elementary function $ln(x)$ is defined over the input domain $(0,
\infty)$. There are four classes of special case inputs:
\begin{small}
  \[
    ln(x) =
    \begin{cases}
      NaN & \text{if } x = NaN \\
      NaN & \text{if } x < 0 \\
      -\infty & \text{if } x = 0 \\
      \infty & \text{if } x = \infty
    \end{cases}
  \]
\end{small}

We use the range reduction technique described in
Appendix~\ref{apx:log} to decompose $x$ into $n$, $F$, and $r$. In the
case of $ln(x)$, the output compensation can be mathematically
computed as,
\begin{small}
  \[
    ln(x) = ln(1 + r) + ln(F) + n \times ln(2)
  \]
\end{small}
To evaluate the output compensation function in double, we
store the correctly rounded value of $ln2 = ln(2)$ in double. We
order the operations in the following way,
\begin{small}
  \[
    ln(x) = (ln(1 + r) + ln(F)) + n \times ln2
  \]
\end{small}
This reduces the amount of numerical error for the inputs $x$ where
the magnitude of $n$ is large. After range reduction, there were
roughly 7.2 million reduced inputs. The reduced inputs are in the
range of $r \in [0, \frac{1}{128})$. We created an approximation of
$ln(1 + r)$ using \tool to generate a piecewise polynomial with
$2^{10}$ polynomials of degree 3.

\subsection{$log_2(x)$}
The elementary function $log_2(x)$ is defined over the input domain $(0,
\infty)$. There are four classes of special case inputs:
\begin{small}
  \[
    log_2(x) =
    \begin{cases}
      NaN & \text{if } x = NaN \\
      NaN & \text{if } x < 0 \\
      -\infty & \text{if } x = 0 \\
      \infty & \text{if } x = \infty
    \end{cases}
  \]
\end{small}

We use the range reduction technique described in
Appendix~\ref{apx:log} to decompose $x$ into $n$, $F$, and $r$. In the
case of $log_2(x)$, the output compensation can be mathematically
computed as,
\begin{small}
  \[
    log_2(x) = log_2(1 + r) + log_2(F) + n
  \]
\end{small}
To evaluate the output compensation function in double, we
order the operations in the following way,
\begin{small}
  \[
    log_2(x) = (log_2(1 + r) + log_2(F)) + n
  \]
\end{small}
This reduces the amount of numerical error for the inputs $x$ where
the magnitude of $n$ is large. After range reduction, there were
roughly 7.2 million reduced inputs. The reduced inputs are in the
range of $r \in [0, \frac{1}{128})$. We created an approximation of
$log_2(x)$ using \tool to generate a piecewise polynomial with $2^{8}$
polynomials of degree 3.

\subsection{$log_{10}(x)$}
The elementary function $log_{10}(x)$ is defined over the input domain
$(0, \infty)$. There are four classes of special case inputs:
\begin{small}
  \[
    log_{10}(x) =
    \begin{cases}
      NaN & \text{if } x = NaN \\
      NaN & \text{if } x < 0 \\
      -\infty & \text{if } x = 0 \\
      \infty & \text{if } x = \infty
    \end{cases}
  \]
\end{small}

We use the range reduction technique described in
Appendix~\ref{apx:log} to decompose $x$ into $n$, $F$, and $r$. In the
case of $log_{10}(x)$, the output compensation can be mathematically
computed as,
\begin{small}
  \[
    log_{10}(x) = log_{10}(1 + r) + log_{10}(F) + n \times log_{10}(2)
  \]
\end{small}
To generate efficient piecewise polynomial approximation, we store the
value of $log_{10}(2)$ in two double values, $t_h$ and $t_l$
such that $t_h + t_l$ is a correctly rounded value of $log_{10}(2)$
with 106 precision bits (each double value has 53 precision
bits). The value $t_h$ stores the higher 53 precision bits and $t_l$
stores the lower 53 precision bits. Then, we evaluate the output
compensation function in double with the following order of
operation:
\begin{small}
  \[
    log_{10}(x) = ((log_{10}(1 + r) + n \times t_l) + log_{10}(F)) + n \times t_h
  \]
\end{small}
This reduces the amount of numerical error for the inputs $x$ where
the magnitude of $n$ is large. After range reduction, there were
roughly 7.2 million reduced inputs. The reduced inputs are in the
range of $r \in [0, \frac{1}{128})$. We created an approximation of
$log_{10}(x)$ using \tool to generate a piecewise polynomial with
$2^{8}$ polynomials of degree 3.

\subsection{$e^x$}
The $e^x$ function is defined over the input domain $(-\infty,
\infty)$. There are 4 classes of special case inputs:
\begin{small}
  \[
    e^{x} =
    \begin{cases}
      NaN & \text{if } x = NaN \\
      0 & \text{if } x \leq -103.97\dots \\
      1.0 & \text{if } -2.98\dots \times 10^{-8} \leq x \leq 5.96\dots \times 10^{-8} \\
      \infty & \text{if } x  \geq 88.72
    \end{cases}
  \]
\end{small}

We use the range reduction technique described in
Appendix~\ref{apx:exp} to perform range reduction and output
compensation. After range reduction, there are roughly 520 million
reduced inputs in the domain
$r \in [-\frac{ln(2)}{64}, \frac{ln(2)}{64}]$. We created an
approximation function for negative values of reduced inputs $r$ and
an approximation function for positive values of reduced inputs to
efficiently split the reduced input domain. Using \tool, we generated
a piecewise polynomial with $2^7$ polynomials of degree 4 for negative
reduced inputs and a piecewise polynomial with $2^7$ polynomials of
degree 4 for positive reduced inputs.

\subsection{$2^x$}
The $2^x$ function is defined over the input domain $(-\infty,
\infty)$. There are 4 classes of special case inputs:
\begin{small}
  \[
    2^{x} =
    \begin{cases}
      NaN & \text{if } x = NaN \\
      0 & \text{if } x \leq -150 \\
      1.0 & \text{if } -4.29\dots \times 10^{-8} \leq x \leq 8.59\dots \times 10^{-8} \\
      \infty & \text{if } x  \geq 128
    \end{cases}
  \]
\end{small}

We use the range reduction technique described in
Appendix~\ref{apx:exp} to perform range reduction and output
compensation. After range reduction, there are roughly 303 million
reduced inputs in the domain $r \in [-\frac{1}{64}, \frac{1}{64}]$. We
created an approximation function for negative values of reduced
inputs $r$ and an approximation function for positive values of
reduced inputs to efficiently split the reduced input domain. Using
\tool, we generated a piecewise polynomial with $2^4$ polynomials of
degree 4 for negative reduced inputs and a piecewise polynomial with
$2^3$ polynomials of degree 4 for positive reduced inputs.

\subsection{$10^x$}
The $10^x$ function is defined over the input domain $(-\infty,
\infty)$. There are 4 classes of special case inputs:
\begin{small}
  \[
    10^{x} =
    \begin{cases}
      NaN & \text{if } x = NaN \\
      0 & \text{if } x \leq -45.15\dots \\
      1.0 & \text{if } -1.29\dots \times 10^{-8} \leq x \leq 2.58\dots
      \times 10^{-8} \\
      \infty & \text{if } x  \geq 38.53\dots
    \end{cases}
  \]
\end{small}

We use the range reduction technique described in
Appendix~\ref{apx:exp} to perform range reduction and output
compensation. After range reduction, there are roughly 521 million
reduced inputs in the domain
$r \in [-\frac{log_{10}(2)}{64}, \frac{log_{10}(2)}{64}]$. We created an
approximation function for negative values of reduced inputs $r$ and
an approximation function for positive values of reduced inputs to
efficiently split the reduced input domain. Using \tool, we generated
a piecewise polynomial with $2^6$ polynomials of degree 4 for negative
reduced inputs and a piecewise polynomial with $2^7$ polynomials of
degree 3 for positive reduced inputs.

\subsection{$Sinpi(x)$}
The $sinpi(x)$ function is defined over the input domain $(-\infty,
\infty)$. There are three classes of special cases:
\begin{small}
\[
  sinpi(x) =
  \begin{cases}
    \pi x & \text{if } |x| < 1.173\dots \times 10^{-7} \\
    0 & \text{if } |x| \geq  2^{23}\\
    NaN & \text{if } x = NaN \text{ or } x = \pm \infty
  \end{cases}
  \]
\end{small}

We use the range reduction technique described in
Appendix~\ref{apx:sinpi} and evaluate the range reduction and output
compensation in double. After range reduction, there were
roughly 117 million reduced inputs in the domain $r \in [0,
\frac{1}{512}]$. The output compensation function for $sinpi(x)$ uses
both $sinpi(r)$ and $cospi(r)$. Similarly, the output compensation
function for $cospi(x)$ uses the approximation of both $sinpi(r)$ and
$cospi(r)$. Thus, we generated an approximation function for
$sinpi(r)$ and an approximation function for $cospi(r)$ that can be
used to compute both $sinpi(x)$ and $cospi(x)$. Using \tool, we
created a single polynomial of degree 5 for $sinpi(r)$ and a single
polynomial of degree 4 for $cospi(r)$.

\subsection{$Cospi(x)$}
The $cospi(x)$ function is defined over the input domain $(-\infty,
\infty)$. There are three classes of special cases:
\begin{small}
\[
  cospi(x) =
  \begin{cases}
    1.0 & \text{if } |x| < 7.771 \times 10^{-5}\\
    (-1)^{(|x| \:mod \: 2)} \times 1.0 & \text{if } |x| \geq 2^{23}\\
    NaN & \text{if } x = NaN \text{ or } x = \pm \infty \\
  \end{cases}
\]
\end{small}

We use the range reduction technique described in
Appendix~\ref{apx:cospi}. We evaluate the range reduction and output
compensation in double. After range reduction, there were
roughly 40 million reduced inputs in the domain $r \in [0,
\frac{1}{512}]$. The output compensation function for $cospi(x)$ uses
both $sinpi(r)$ and $cospi(r)$. These approximation functions are also
used in $sinpi(x)$. Thus, we generated approximation functions for
$sinpi(r)$ and $cospi(r)$ that can be used to compute the results for
both $sinpi(x)$ and $cospi(x)$. Using \tool, we
created a single polynomial of degree 5 for $sinpi(r)$ and a single
polynomial of degree 4 for $cospi(r)$.

\subsection{$Sinh(x)$}
The $sinh(x)$ function is defined over the input domain $(-\infty,
\infty)$. There are four classes of special cases:
\[
  \text{sinh(x)} =
  \begin{cases}
    -\infty & \text{if } x <= -89.415\dots \nonumber  \\ 
    x & \text{if } |x| \leq 4.436\dots \times 10^{-4} \nonumber\\ 
    \infty & \text{if }  x >= 89.415\dots \nonumber\\ 
    NaN & \text{if } x = NaN \nonumber
  \end{cases}
\]

We use the range reduction technique described in
Appendix~\ref{apx:sinh} and evaluate the range reduction and output
compensation functions in double. After range reduction,
there were roughly 147 million reduced inputs in the domain
$r \in [0, \frac{ln(2)}{64})$. Because the output compensation
function for both $sinh(x)$ and $cosh(x)$ uses approximations of
$sinh(r)$ and $cosh(r)$, we generated a piecewise polynomial for
$sinh(r)$ and a piecewise polynomial for $cosh(r)$ that can be used to
compute both $sinh(x)$ and $cosh(x)$ correctly. Using \tool, we
created a piecewise polynomial with $2^6$ polynomials of degree 5 for
$sinh(r)$ and a piecewise polynomial with $2^6$ polynomials of degree
4 for $cosh(r)$.

\subsection{$Cosh(x)$}
The $cosh(x)$ functions is defined over the input domain $(-\infty,
\infty)$. There are three classes of special cases:
\[
  \text{cosh(x)} =
  \begin{cases}
    1.0 & \text{if } |x| \leq 3.452\dots \times 10^{-4} \\
    \infty & \text{if }  |x| >= 89.415\dots \\
    NaN & \text{if } x = NaN
  \end{cases}
\]

We use the range reduction technique described in
Appendix~\ref{apx:cosh} and evaluate the range reduction and output
compensation functions in double. After range reduction,
there were roughly 151 million reduced inputs in the domain. We
approximate $sinh(r)$ and $cosh(r)$ for the reduced inputs in the
domain $r \in [0, \frac{ln(2)}{64})$. Since the output compensation
function of $sinh(x)$ also uses approximations of $sinh(r)$ and
$cosh(r)$, we generate a piecewise polynomial for $sinh(r)$ and a
piecewise polynomial for $cosh(r)$ that can be used for the output
compensation of both $sinh(x)$ and $cosh(x)$. Using \tool, we created
a piecewise polynomial with $2^6$ polynomials of degree 5 for
$sinh(r)$ and a piecewise polynomial with $2^6$ polynomials of degree
4 for $cosh(r)$.

\section{32-bit Posit32 Functions}
\label{apx:posit}

In this section, we explain the 32-bit posit
(posit32) functions in \tool. We describe the special cases
and any specific details on the range reduction technique used for
each function.

\subsection{$ln(x)$}
The elementary function $ln(x)$ is defined over the input domain $(0,
\infty)$. There are two classes of special case inputs of $ln(x)$ for
posit32:
\begin{small}
  \[
    ln(x) =
    \begin{cases}
      NaR & \text{if } x \leq 0 \\
      NaR & \text{if } x = NaR
    \end{cases} 
  \]
\end{small}

We use the range reduction technique described in
Appendix~\ref{apx:log} to decompose $x$ into $n$, $F$, and $r$. In the
case of $ln(x)$, the output compensation can be mathematically
computed as,
\begin{small}
  \[
    ln(x) = ln(1 + r) + ln(F) + n \times ln(2)
  \]
\end{small}
To compute $n \times ln(2)$ accurately, we store the value of $ln(2)$
in two double values, $t_h$ and $t_l$, such that $t_h + t_l$
stores the correctly rounded value of $ln(2)$ with 106 precision bits
(double type has 53 precision bits). The value $t_h$ stores
the higher 53 precision bits and $t_l$ stores the lower 53 precision
bits. We evaluate the output compensation function in double
with the following order,
\begin{small}
  \[
    ln(x) = ((ln(1 + r) + n \times t_l) + ln(F)) + n \times t_h
  \]
\end{small}
This reduces the amount of numerical error for the inputs $x$ where
the magnitude of $n$ is large. After range reduction, there were
roughly 115 million reduced inputs. The reduced inputs are in the
range of $r \in [0, \frac{1}{128})$. We created an approximation of
$ln(1 + r)$ using \tool to generate a piecewise polynomial with
$2^{10}$ polynomials of degree 4.

\subsection{$log_2(x)$}
The elementary function $log_2(x)$ is defined over the input domain $(0,
\infty)$. There are two classes of special case inputs of $log_2(x)$ for
posit32:
\begin{small}
  \[
    log_2(x) =
    \begin{cases}
      NaR & \text{if } x \leq 0 \\
      NaR & \text{if } x = NaR
    \end{cases} 
  \]
\end{small}

We use the range reduction technique described in
Appendix~\ref{apx:log} to decompose $x$ into $n$, $F$, and $r$. The
output compensation function for $log_2(x)$ can be mathematically
computed as,
\begin{small}
  \[
    log_2(x) = log_2(1 + r) + log_2(F) + n
  \]
\end{small}

The output compensation function that we use for $log_a(x)$
experiences cancellation error when $x = 1 - \epsilon$ for small
values of $\epsilon$. In such cases, $log_2(x) \approx 0$, $m = -1$,
and $log_2(1 + r) + log_2(F) \approx 1$. The numerical error caused by
cancellation error in the output compensation function of $log_2(x)$
for posit32 poses a challenge in generating piecewise
polynomials of reasonable degree and size that satisfies all reduced
input and interval constraints.

Thus, we evaluate the output compensation function in double
with the following order of operations,
\begin{small}
  \[
    log_2(x) = (log_2(1 + r) + n) + log_2(F)
  \]
\end{small}
This order subtracts two values with largest difference in magnitude,
before subtracting two values with similar magnitude. After range
reduction, there were roughly 115 million reduced inputs. The reduced
inputs are in the range of $r \in [0, \frac{1}{128})$. We created an
approximation of $log_2(1 + r)$ using \tool to generate a piecewise
polynomial with $2^{8}$ polynomials of degree 4. Comparatively, when
we generated a piecewise polynomial for the output compensation
function that evaluates in the following order,
\begin{small}
  \[
    log_2(x) = (log_2(1 + r) + log_2(F)) + n
  \]
\end{small}
\tool generated a piecewise polynomial with $2^{11}$ polynomials of
degree 4. In all other cases of $log_a(x)$ for float or
posit32, adding $log_a(1 + r) + n \times log_a(2)$, then
adding $log_a(F)$ at the end did not produce piecewise polynomial with
smaller number of polynomials.

\subsection{$log_{10}(x)$}
The elementary function $log_{10}(x)$ is defined over the input domain
$(0, \infty)$. There are two classes of special case inputs of
$log_{10}(x)$ for posit32:
\begin{small}
  \[
    log_{10}(x) =
    \begin{cases}
      NaR & \text{if } x \leq 0 \\
      NaR & \text{if } x = NaR
    \end{cases} 
  \]
\end{small}

We use the range reduction technique described in
Appendix~\ref{apx:log} to decompose $x$ into $n$, $F$, and $r$. The
output compensation function of $log_{10}(x)$ can be mathematically
computed as,
\begin{small}
  \[
    log_{10}(x) = log_{10}(1 + r) + log_{10}(F) + n \times log_{10}(2)
  \]
\end{small}
To compute $n \times log_{10}(2)$ accurately, we store $log_{10}(2)$
in two double values, $t_h$ and $t_l$. The sum $t_h + t_l$,
if evaluated in real numbers, is the correctly rounded value of
$log_{10}(2)$ with 106 precision bits. The value $t_h$ stores the
higher 53 precision bits and $t_l$ stores the lower 53 precision
bits. We evaluate the output compensation function in double
in the following order,
\begin{small}
  \[
    log_{10}(x) = ((log_{10}(1 + r) + n \times t_l) + log_{10}(F)) + n \times t_h
  \]
\end{small}
After range reduction, there were roughly 115 million reduced
inputs. The reduced inputs are in the range of
$r \in [0, \frac{1}{128})$. We created an approximation of $log_{10}(1 + r)$
using \tool to generate a piecewise polynomial with $2^{12}$
polynomials of degree 4.

\subsection{$e^x$}
The $e^x$ function is defined over the input domain $(-\infty,
\infty)$. There are 4 classes of special case inputs:
\begin{small}
  \[
    e^{x} =
    \begin{cases}
      NaR & \text{if } x = NaR \\
      2^{-120} & \text{if } x \leq -81.7\dots \\
      1.0 & \text{if } -1.86\dots \times 10^{-9} \leq x \leq 3.72\dots \times 10^{-9} \\
      2^{120} & \text{if } x  \geq 81.7\dots
    \end{cases}
  \]
\end{small}

We use a range reduction technique similar to the technique described
in Appendix~\ref{apx:exp} to perform range reduction and output
compensation. More specifically, we split $x$ into $128$ segments
instead of $64$ segments:
\[
  x = nln(2) + \frac{j}{128}ln(2) + r
\]
where $n$ is an integer, $j$ is a value in a set $\{0, 1, 2, \dots,
127\}$ and $|r| \leq \frac{ln(2)}{128}$.

Then, the output compensation formula is adjusted accordingly:
\begin{small}
\begin{align}
  e^x = e^{nln(2) + \frac{j}{128}ln(2) + r} &= e^{nln(2)} \times
                                             e^{\frac{j}{64}ln(2)} \times e^{r} \nonumber \\
  &= 2^n \times 2^{\frac{j}{128}} \times e^{r} \nonumber
\end{align}
\end{small}
We pre-compute and store the value of $2^{\frac{j}{128}}$ in a table
(\ie 128 values in total) and approximate $e^r$ for the input domain
of $r \in [-\frac{ln(2)}{128}, \frac{ln(2)}{128}]$.

After range reduction, there are roughly 3.5 billion reduced inputs in
the domain $r \in [-\frac{ln(2)}{128}, \frac{ln(2)}{128}]$. We created
an approximation function for negative values of reduced inputs $r$
and an approximation function for positive values of reduced inputs to
efficiently split the reduced input domain. Using \tool, we generated
a piecewise polynomial with $2^{12}$ polynomials of degree 3 for negative
reduced inputs and a piecewise polynomial with $2^{12}$ polynomials of
degree 3 for positive reduced inputs.

\subsection{$2^x$}
The $2^x$ function is defined over the input domain $(-\infty,
\infty)$. There are 4 classes of special case inputs:
\begin{small}
  \[
    2^{x} =
    \begin{cases}
      NaR & \text{if } x = NaR \\
      2^{-120} & \text{if } x \leq -118\dots \\
      1.0 & \text{if } -2.68\dots \times 10^{-9} \leq x \leq 5.37\dots \times 10^{-9} \\
      2^{120} & \text{if } x  \geq 118\dots
    \end{cases}
  \]
\end{small}

We use the range reduction technique described in
Appendix~\ref{apx:exp} to perform range reduction and output
compensation. After range reduction, there are roughly 790 million
reduced inputs in the domain $r \in [-\frac{1}{64}, \frac{1}{64}]$. We
created an approximation function for negative values of reduced
inputs $r$ and an approximation function for positive values of
reduced inputs to efficiently split the reduced input domain. Using
\tool, we generated a piecewise polynomial with $2^{10}$ polynomials of
degree 3 for negative reduced inputs and a piecewise polynomial with
$2^{12}$ polynomials of degree 3 for positive reduced inputs.

\subsection{$10^x$}
The $10^x$ function is defined over the input domain $(-\infty,
\infty)$. There are 4 classes of special case inputs:
\begin{small}
  \[
    10^{x} =
    \begin{cases}
      NaR & \text{if } x = NaR \\
      2^{-120} & \text{if } x \leq -35.5\dots \\
      1.0 & \text{if } -8.08\dots \times 10^{-10} \leq x \leq 1.61\dots \times 10^{-9} \\
      2^{120} & \text{if } x  \geq 35.5\dots
    \end{cases}
  \]
\end{small}

We use the range reduction technique described in
Appendix~\ref{apx:exp} to perform range reduction and output
compensation. After range reduction, there are roughly 3.4 billion
reduced inputs in the domain
$r \in [-\frac{log_{10}(2)}{64}, \frac{log_{10}(2)}{64}]$. We created an
approximation function for negative values of reduced inputs $r$ and
an approximation function for positive values of reduced inputs to
efficiently split the reduced input domain. Using \tool, we generated
a piecewise polynomial with $2^{13}$ polynomials of degree 3 for negative
reduced inputs and a piecewise polynomial with $2^{13}$ polynomials of
degree 3 for positive reduced inputs.

\subsection{$Sinh(x)$}
The $sinh(x)$ function is defined over the input domain $(-\infty,
\infty)$. There are four classes of special cases:
\[
  \text{sinh(x)} =
  \begin{cases}
    -2^{120} & \text{if } x <= -82.4\dots \nonumber  \\ 
    x & \text{if } |x| \leq 2.79\dots \times 10^{-4} \nonumber\\ 
    2^{120} & \text{if }  x >= 82.4\dots \nonumber\\ 
    NaR & \text{if } x = NaR \nonumber
  \end{cases}
\]

We use the range reduction technique described in
Appendix~\ref{apx:sinh} and evaluate the range reduction and output
compensation functions in double. After range reduction,
there were roughly 1.6 billion reduced inputs in the domain
$r \in [0, \frac{ln(2)}{64})$. Because the output compensation
function for $sinh(x)$ uses approximations of
$sinh(r)$ and $cosh(r)$, we generated a piecewise polynomial for
$sinh(r)$ and a piecewise polynomial for $cosh(r)$. Using \tool, we
created a piecewise polynomial with $2^{14}$ polynomials of degree 5 for
$sinh(r)$ and a piecewise polynomial with $2^{14}$ polynomials of degree
4 for $cosh(r)$.

\subsection{$Cosh(x)$}
The $cosh(x)$ functions is defined over the input domain $(-\infty,
\infty)$. There are three classes of special cases:
\[
  \text{cosh(x)} =
  \begin{cases}
    1.0 & \text{if } |x| \leq 8.63\dots \times 10^{-5} \\
    2^{120} & \text{if }  |x| >= 82.4\dots \\
    NaR & \text{if } x = NaR
  \end{cases}
\]

We use the range reduction technique described in
Appendix~\ref{apx:cosh} and evaluate the range reduction and output
compensation functions in double. After range reduction,
there were roughly 1.7 billion reduced inputs in the domain. We
approximate $sinh(r)$ and $cosh(r)$ for the reduced inputs in the
domain $r \in [0, \frac{ln(2)}{64})$. Since the output compensation
function of $cosh(x)$ uses approximations of $sinh(r)$ and
$cosh(r)$, we generate a piecewise polynomial for $sinh(r)$ and a
piecewise polynomial for $cosh(r)$. Using \tool, we created
a piecewise polynomial with $2^{14}$ polynomials of degree 3 for
$sinh(r)$ and a piecewise polynomial with $2^{12}$ polynomials of degree
6 for $cosh(r)$.

\end{document}